\documentclass[final,3p,times,preprint,11pt]{elsarticle}

\usepackage{tabularx}
\usepackage{amssymb}

\usepackage{graphicx}
\usepackage{subfigure}

\usepackage{caption}

\DeclareCaptionLabelFormat{boldfigure}{\textbf{Fig. #2}}
\DeclareCaptionLabelSeparator{boldsep}{\textbf{. }}
\usepackage{amsmath}
\usepackage{bm}
\usepackage{booktabs} 
\usepackage{diagbox} 
\usepackage{multirow} 

\usepackage[section]{placeins}

\begin{document}

\begin{frontmatter}



\title{Physics-informed neural networks for unsteady incompressible flows with time-dependent moving boundaries}


\author[inst1]{Yongzheng Zhu}

\author[inst2]{Weizhen Kong}

\author[inst1]{Jian Deng}

\author[inst1]{Xin Bian\corref{cor1}}
\ead{bianx@zju.edu.cn}

\affiliation[inst1]{organization={State Key Laboratory of Fluid Power and Mechatronic Systems, Department of Engineering Mechanics, Zhejiang University},
            city={Hangzhou},
            postcode={310027}, 
            country={China}}
\affiliation[inst2]{organization={China Ship Scientific Research Center},
            city={Wuxi},
            postcode={214082}, 
            country={China}}
\cortext[cor1]{Corresponding author.}

\begin{abstract}
Physics-informed neural networks (PINNs) employed in fluid mechanics deal primarily with stationary boundaries. This hinders the capability to address a wide range of flow problems involving moving bodies. 
To this end, we propose a novel extension, which enables PINNs to solve incompressible flows with time-dependent moving boundaries. 
More specifically, we impose Dirichlet constraints of velocity at the moving interfaces and define new loss functions for the corresponding training points.
Moreover, we refine training points for flows around the moving boundaries for accuracy. 
This effectively enforces the no-slip condition of the moving boundaries.
With an initial condition, the extended PINNs solve unsteady flow problems with time-dependent moving boundaries and still have the flexibility to leverage partial data to reconstruct the entire flow field.
Therefore, the extended version inherits the amalgamation of both physics and data from the original PINNs.
With a series of typical flow problems, we demonstrate the effectiveness and accuracy of the extended PINNs.
The proposed concept allows for solving inverse problems as well, which calls for further investigations.
\end{abstract}




\end{frontmatter}



\section{Introduction}
\label{sec:Introduction}
In recent years, machine learning has achieved enormous accomplishments in part due to the technological innovation in graph processing units and healthy development of software ecosystems. It finds applications in a variety of disciplines, including physics, linguistics, biology, and many others, as well as in scenarios such as image recognition, natural language processing, and autonomous driving~\citep{lecun2015deep}. Inspired by these developments, applied mathematicians have swiftly proposed various ingenious frameworks based on machine learning, aiming to shift the traditional paradigm of computational science and engineering~\citep{brunton2020machine, karniadakis2021physics}. The novel frameworks may be data-driven, physics-informed, partial differential equations (PDEs)-regularized, in a variational form, optimization of discrete loss and so on~\citep{Brunton2016, sirignano2018dgm, E2018, raissi2019physics,Karnakov2022}. 
In particular, the physics-informed neural networks (PINNs) have stimulated a wide range of research efforts~\citep{karniadakis2021physics}. The key idea behind PINNs is to incorporate the residual of PDEs into the loss function of the neural network and leverage its powerful nonlinear fitting abilities to approximate the solutions~\citep{raissi2019physics}. Given the impressive performance, researchers have extended PINNs to solve various types of differential equations~\citep{pang2019fpinns,fang2019physics,zhang2020learning}. 
Many algorithmic aspects have also been considered to facilitate the training of PINNs to improve the accuracy and/or convergence. These include but are not limited to adaptive activation functions~\citep{jagtap2020adaptive}, adaptive weights~\citep{wang2021understanding}, gradient-enhanced approaches~\citep{yu2022gradient}, artificial-viscosity regularized loss~\citep{he2023}, residual-based and/or solution-gradient based adaptive sampling~\citep{wu2023comprehensive, Mao2023}, among others. Furthermore, PINNs are also extended to deal with multi-fidelity data and multiscale physics~\citep{Meng2020, Meng2020a}. To tackle various differential equations steadily, an open-source code known as DeepXDE has been provided as a platform for streamlined training of PINNs~\citep{lu2021deepxde}.

Most known physics laws are described by PDEs and therefore, PINNs provides a new paradigm for solving problems in various fields. Moreover, the solution strategy is universal for both forward and inverse problems, whether with partial data or without data. In fluid mechanics, Navier-Stokes (NS) equations are regarded as the fundamental PDEs. Raissi et al.~\citep{raissi2019physics} applied PINNs to infer an unknown coefficient of the NS equations from flow data around a stationary cylinder. Furthermore, they adjusted PINNs for flow visualization, where concentration fields were observed from artwork or medical images to reconstruct the complete flow fields~\citep{raissi2020hidden}. Rao et al.~\citep{rao2020physics} utilized PINNs as a direct numerical simulation~(DNS) tool to study flow around a stationary cylinder at low Reynolds number. Jin et al.~\citep{jin2021nsfnets} evaluated the accuracy, convergence, and computational cost of PINNs on laminar flows such as flow around a cylinder and turbulent flows in pipeline using DNS dataset as benchmark. Wang et al.~\citep{wang2021deep} proposed an algorithm to enable PINNs to deal with 1D two-phase Stefan problems with free boundaries. Cai et al.~\citep{cai2021physics} explored PINNs for heat transfer problems, in which they utilized sparse measuring points of temperature to infer both temperature and velocity fields of the entire domain. Despite so many successful applications of PINNs in fluid mechanics, flows with moving boundaries are rarely considered, which are frequently encountered in nature and industry, such as flow over a moving blunt body, bird/insect hovering and fish swimming~\citep{wang2000two, wang2000vortex, wu2011fish, bian2014hydrodynamic, Verma2018, Chatzimanolakis2022}. One exception is the work of Raissi et al.~\citep{raissi2019deep}, where they adopted a transformed coordinate attached to the cylinder to resolve the flow for vortex-induced vibration. The coordinate transformation is elegant, but it is not suitable for general moving boundaries such as multiple blunt bodies in flows. In contrast, computational fluid dynamics (CFD) methods have been employed routinely to solve flow problems with moving boundaries~\citep{moukalled2016finite, peyret2002spectral, karniadakis2005spectral}, although they may require complex meshing techniques such as morphing mesh~\citep{biancolini2014sails} and overset mesh~\citep{jarkowski2014towards}, or interpolation techniques such as the immersed boundary method (IBM)~\citep{peskin2002immersed}. To bridge this gap and allow PINNs to handle flow problems with moving boundaries, we introduce new loss functions proper for training points at moving interfaces and around moving boundaries in the entire temporal domain. Therefore, we are still able to leverage the power and flexibility of the classical PINNs to infer all flow fields, whether with partial data or without any exogenous data. Interactions between fluid and a moving body may be categorized into two types: one-way coupling, where the motion of the body is prescribed and its influence on the fluid dynamics needs to be computed; two-way coupling, where interface between the fluid and the body is part of the solution itself and dynamics of both parts must be computed as well. In this work, we focus on the first type, as in reality it is rather straightforward to extract the trajectories of blunt bodies and reconstruction of the flow fields around moving bodies often remain obscure.

This rest of the paper is organized as follows. In Section~\ref{sec:Overview of PINNs}, we review PINNs briefly and present the no-slip velocity condition for time-dependent moving bodies in detail in Section~\ref{sec:Strategy for handling moving boundaries}. Direct numerical simulation results of PINNs are validated by three examples in Section~\ref{sec:Functioning as a DNS solver}. In Section~\ref{sec:Reconstructing the flow field}, we further reconstruct the whole flow fields with partial data in three representative cases. Discussions and conclusions are made in Section~\ref{sec:Discussions and conclusions}. Auxiliary data from CFD methods are presented in the Appendix.

\section{Physics-informed neural networks}
\label{sec:Overview of PINNs}
Consider the parametric nonlinear PDEs in a general form expressed as follows:
\begin{equation}
f\left( {{\bm x};\frac{{\partial s}}{{\partial {x_1}}}, \ldots ,\frac{{\partial s}}{{\partial {x_d}}};\frac{{{\partial ^2}s}}{{\partial {x_1}\partial {x_1}}}, \ldots ,\frac{{{\partial ^2}s}}{{\partial {x_1}\partial {x_d}}}; \ldots ; {\lambda}} \right) = 0, \quad {\bm x} \in \Omega,
\end{equation}
with the boundary condition and initial condition:
\begin{eqnarray}
&& {\cal B}(s,{\bm x}) = 0,\quad {\bm x} \in \partial \Omega,
\\
&& {\cal I}(s,{\bm x}) = 0,\quad {\bm x} \in {{\Gamma}_i}.
\end{eqnarray}
Here ${\bm x}=\left[x_1, x_2, \ldots, x_d\right]$ are the independent variables, $f$ denotes the linear or nonlinear differential operators, $s$ is the solution, and $\lambda=\left[\lambda_1, \lambda_2, \ldots\right]$ are the parameters for combining each component. $\Omega$ and $\partial \Omega$ represent the computational domain and the boundaries, respectively. ${\Gamma}_i$ stands for the space of ${\bm x}$ at the initial snapshot.

Within the framework of PINNs~\citep{raissi2019physics}, a fully connected neural network (FCNN) is exploited as a highly nonlinear function $\hat s({\bm x})$ to approximate the solution $s$ of the PDEs. It is composed of an input layer, multiple hidden layers, and an output layer,
while each layer contains several neurons with weights, biases, and non-linear activation function $\sigma(\cdot)$. Let ${\bm x}$ be the input and the implicit variable of the $i$th hidden layer be ${\bm y}^i$, then a FCNN with $L$ layers can be expressed as:
\begin{equation}
\left\{
\begin{aligned}
{\bm y}^0& = {\bm x}, \\
{\bm y}^i& = \sigma\left(\bm{W}^i {\bm y}^{i-1}+\bm{b}^i\right), \quad 1 \leq i \leq L-1, \\
{\bm y}^i& = \bm{W}^i {\bm y}^{i-1}+\bm{b}^i, \quad i=L,
\end{aligned}
\right.
\end{equation}
where $\bm{W}^i$ denotes the weight matrix and $\bm{b}^i$ is the bias vector to be trained at $i$th layer, respectively. 
$\sigma(\cdot)$ functions element-wise. 

Since $\hat s$ is the approximate solution,
each component in the PDEs can be obtained by taking derivatives of $\hat s$ with respect to $x_i$ one or more times by the automatic differentiation (AD) techniques combined with the chain rule~\citep{rall1981automatic}. 
Subsequently, we define a composite loss function as the sum of the residuals of the equation, boundary condition, initial condition, and labeled data:
\begin{equation}
{\cal L}({\cal T}) = {w_f}{{\cal L}_f}\left({\cal T}_f \right) + {w_b}{{\cal L}_b}\left({\cal T}_b \right) + {w_i}{{\cal L}_i}\left( {\cal T}_i \right)+ {w_d}{{\cal L}_{data}}\left({\cal T}_{data} \right),
\label{eq:totalloss}
\end{equation}
where $w_f$, $w_b$, $w_i$ and $w_d$ are the corresponding weights
and ${\cal T}_f$, ${\cal T}_b$, ${\cal T}_i$ and ${\cal T}_{data}$ are the corresponding sets of training points, respectively.

With the target of minimizing the loss function, the neural network is trained by optimizing the weights and biases through the back-propagation algorithm with an optimizer such as Adam~\citep{kingma2014adam}.
The flexibility of PINNs lies in the capability of switching between supervised, weakly supervised, and unsupervised learning approaches. For a supervised learning or data-driven approach, the loss function is guided only by a labeled dataset, that is, ${\cal T}_{data}$,
and PINNs degenerates to be FCNNs.
For an unsupervised learning, the loss function is defined by the equation, the boundary condition, and the initial condition. The training dataset ${\cal T}_f$ corresponds to collocation points, where solutions need to be inferred,
and they are constrained by the points within ${\cal T}_b$ and ${\cal T}_i$.
More frequently, PINNs are employed with all the four types of loss functions of training points.

\section{PINNs for incompressible flows with time-dependent moving boundaries}
\label{sec:Strategy for handling moving boundaries}
\subsection{Governing equations}
The NS equations are expressed as continuity and momentum equations in two dimensional coordinates ${\bm x}=(x, y)$ as follows:
\begin{equation}
\label{eq:NS}
\begin{split}
    \nabla  \cdot {\bm u}& = 0,
    \\
    \frac{{\partial {\bm u}}}{{\partial t}} + ({\bm u} \cdot \nabla ){\bm u}& =  - \nabla p + \frac{1}{{Re}}{\nabla ^2}{\bm u},
\end{split}
\end{equation}
where $t$ is the time, ${\bm u}=\left(u, v\right)$ is the velocity vector, $p$ is the pressure, and $Re=UD/v$ is the Reynolds number defined by the reference velocity $U$, the characteristic length $D$, and the kinematic viscosity $v$. 

If there are solid or truncated boundaries at stationary, one or more constraints for velocity and/or pressure may be specified as follows:
\begin{eqnarray}
\label{Dirichlet_velocity_for_stationary_boundaries}
{\bm u} = {{\bm u}_\Gamma }({\bm x}),\quad {\bm x} \in {\Gamma _D},
\\
p = {p_\Gamma }({\bm x}),\quad {\bm x} \in {\Gamma _D},
\\
\frac{{\partial {\bm u}}}{{\partial {\bm n}}} = 0,\quad {\bm x} \in {\Gamma _N},
\\
\frac{{\partial p}}{{\partial {\bm n}}} = 0,\quad {\bm x} \in {\Gamma _N},
\end{eqnarray}
where ${\bm n}$ is the normal vector at the boundary; ${\Gamma _D}$ and ${\Gamma _N}$ denote the Dirichlet and Neumann boundaries, respectively.

\subsection{Boundary conditions of moving bodies}

\begin{figure}[htb]
    \centering
    \includegraphics[width=0.9\linewidth]{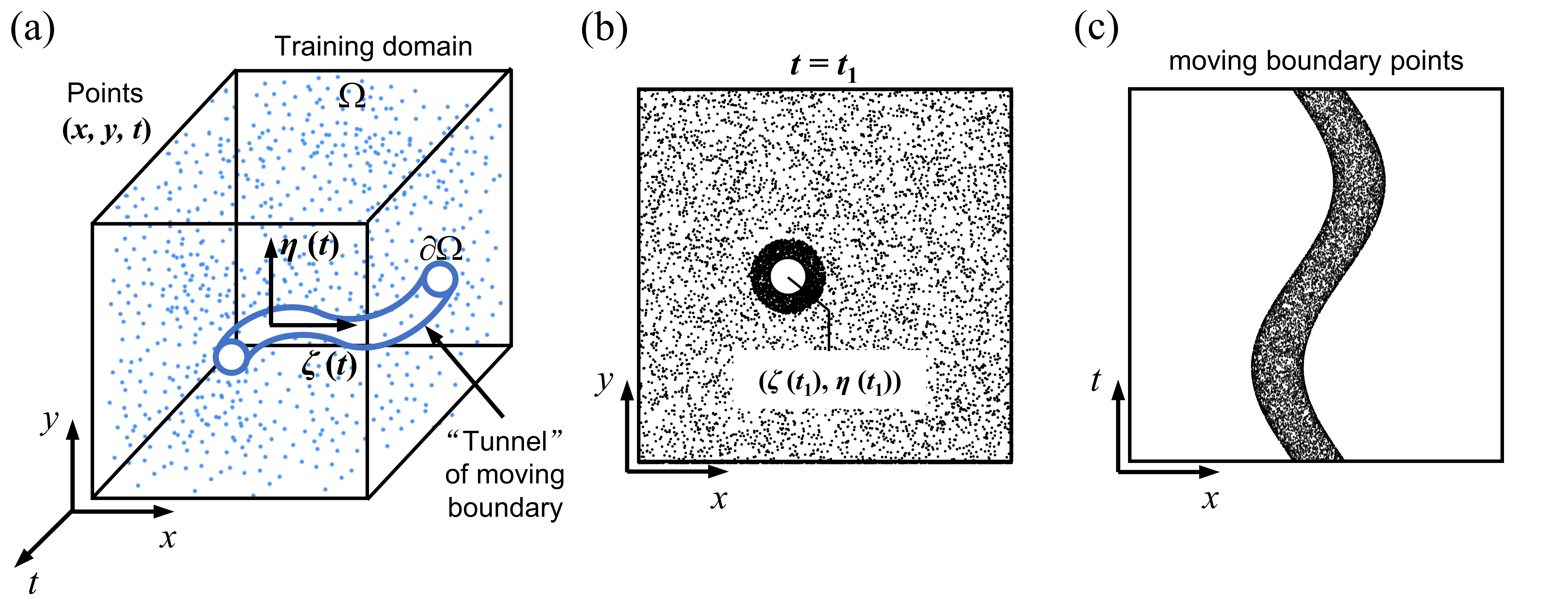}
    \caption{Schematic of a moving cylinder in spatial-temporal domain and its corresponding training points. (a) Training points are randomly distributed in $\Omega$, which excludes the "tunnel" occupied by the travelling cylinder with $\partial \Omega$ as the interface. (b) Distribution of training points in $x-y$ plane at time $t=t_1$, with finer points sampled at the interface and around the cylinder. (c) Aerial view in the y-direction of all training points sampled at the moving interface.}
    \label{fig:moving_boundaries}
\end{figure}
Firstly, we take a cylinder that travels through the fluid constantly with time, as an example. 
Other moving solid bodies can be accomplished similarly.
As the interface between the fluid and solid is time-dependent,
we may "excavate a tunnel" occupied by the cylinder in the spatial-temporal domain, as illustrated by Fig.~\ref{fig:moving_boundaries}(a).
We denote the fluid domain outside of the tunnel as $\Omega$.
Accordingly, an interface $\partial \Omega$ is formed as the cylinder travels through the fluid and it is given as a function of time:
\begin{equation}
\partial \Omega : = f(\bm{\delta}(t), \alpha (t), t),
\end{equation}
where $\bm{\delta} = (\zeta, \eta)$ is the spatial coordinate for the center of the cylinder. 
Moreover, $\alpha$ is its rotation angle vector ,
which shall be useful for a general-shaped moving body.
The Dirichlet velocity constraint according to the no-slip condition is as follows:
\begin{equation}
\label{eq:moving boundary velocity}
{\bm u} = \frac{{\partial \bm{\delta} }}{{\partial t}} + {\bm r} \times \frac{{\partial \alpha }}{{\partial t}},\quad ({\bm x}, t) \in \partial \Omega,
\end{equation}
where $\left({\bm x},t\right)=\left(x, y, t\right)$ is a spatial-temporal coordinate at the interface and ${\bm r}$ is the corresponding radial vector to the center of the cylinder. 

Traditional CFD methods may address this relatively complex issue, for example, by regenerating computational mesh frequently. In contrast, PINNs does not have a time-stepping, but instead optimizes the flow prediction in a continuous spatial-temporal domain via minimizing the composite loss of training points. 
It is simple to see that {\it the time-dependent moving boundary problem in the two-dimensional coordinate of space corresponds to a stationary problem in
the three dimensional coordinate of space and time.}
In this way, one or more moving boundaries can be dealt with steadily. 
Therefore, we firstly generate sufficient training points randomly in the entire $\Omega$, as illustrated in Fig.~\ref{fig:moving_boundaries}(a).
Furthermore, to take into account the stiff flow variation in the vicinity of the moving boundary, we distribute finer training points around the moving boundary, as shown in Fig.~\ref{fig:moving_boundaries}(b). In Fig.~\ref{fig:moving_boundaries}(c), an aerial view of the training points that are sampled at the moving interface, are looked upon in the $y$ direction. Meanwhile, additional training points at the interface $\partial \Omega$ are imposed with Dirichlet velocity constraint.

Theoretically, this strategy can be extended to three spatial dimensions as well,
which just needs to be elevated into four dimensional coordinate of space and time. 
Next we shall introduce the technical details of PINNs' architecture and elaborate each component of the composite loss function.

\subsection{Each loss function and technical details for PINNs}
\begin{figure}[htb]
    \centering
    \includegraphics[width=1.0\linewidth]{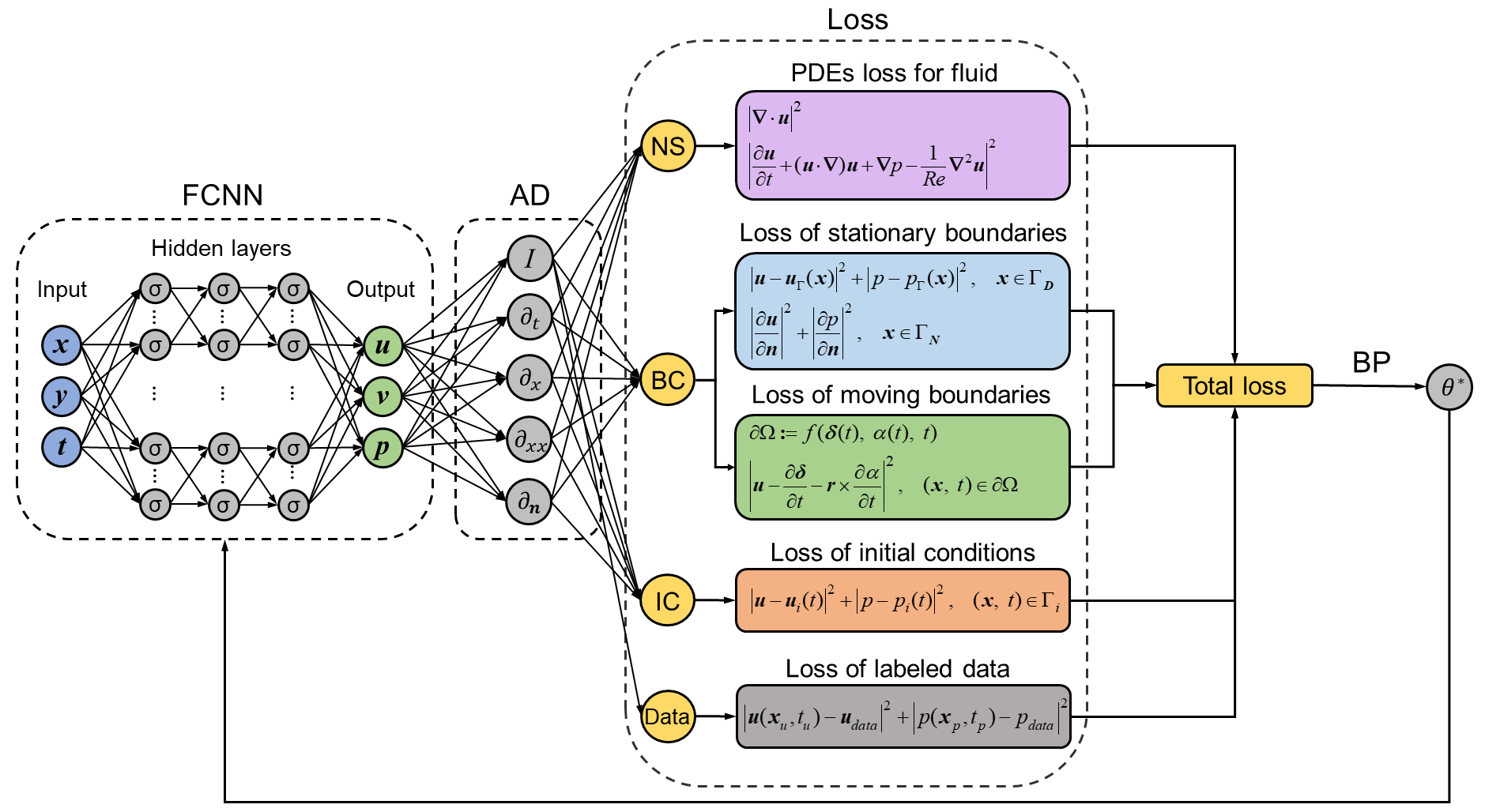}
    \caption{Schematic of PINNs for solving unsteady flow problems with moving boundaries: four types of loss functions and the one for boundary condition consists of stationary and moving parts.}
    \label{fig:FCNN_Architecture}
\end{figure}
Based on the principles of PINNs in Section~\ref{sec:Overview of PINNs},
we take spatial and temporal coordinates $(x,y,t)$ as inputs 
and $(u,v,p)$ as outputs of the neural networks to predict the unsteady velocity and pressure fields.
Fig.~\ref{fig:FCNN_Architecture} illustrates a schematic of PINNs for solving unsteady flow problems with moving boundaries.
The derivatives of $u$, $v$, and $p$ with respect to the inputs are calculated using the AD. 
The composite loss of PINNs is repeated as follows:
\begin{equation}
{\cal L}({\bf{\theta }}) = {w_f}{{\cal L}_f}({\bf{\theta }}) + {w^s_{b}}{{\cal L}^s_{b}}({\bf{\theta }}) + {w^m_{b}}{{\cal L}^m_{b}}({\bf{\theta }}) + {w_i}{{\cal L}_i}({\bf{\theta }}) + {w_{data}}{{\cal L}_{data}}({\bf{\theta }}).
\label{eq:totalloss2}
\end{equation}
The difference from Eq.~(\ref{eq:totalloss}) is that we replace $w_{b}{\cal L}_b$
with $w^s_{b}{\cal L}^s_b + w^m_{b}{\cal L}^m_b$ to differentiate the sub-loss for stationary and moving boundaries, respectively.
Morever, we adopt the Glorot scheme to randomly initialize all weights and biases denoted by 
$\theta$~\citep{glorot2010understanding},
which are to be optimized.
The definitions for all components in Eq.~(\ref{eq:totalloss2}) are expressed in mean squared errors (MSE) as:
\begin{eqnarray}
&&{{\cal L}_f} = \frac{1}{{{N_f}}}\sum\limits_{({\bm x}, t) \in \Omega } {\left\| {\nabla \cdot {\bm u}} \right\|_2^2} + \frac{1}{{{N_f}}}\sum\limits_{({\bm x}, t) \in \Omega } {\left\| {\frac{{\partial {\bm u}}}{{\partial t}} + ({\bm u} \cdot \nabla ){\bm u} + \nabla p - \frac{1}{{Re}}{\nabla ^2}{\bm u}} \right\|_2^2}, 
\\
&&{{\cal L}^s_{b}} = \frac{1}{N^s_b}\sum\limits_{{\bm x} \in {\Gamma _{D}}} {\left( {\left\| {{\bm u} - {{\bm u}_\Gamma }({\bm x})} \right\|_2^2 + \left\| {p - {p_\Gamma }({\bm x})} \right\|_2^2} \right)} + \frac{1}{{{N^s_{b}}}}\sum\limits_{x \in {\Gamma _{N}}} {\left( {\left\| {\frac{{\partial {\bm u}}}{{\partial {\bm n}}}} \right\|_2^2 + \left\| {\frac{{\partial p}}{{\partial {\bm n}}}} \right\|_2^2} \right)}, 
\\
&&{{\cal L}^m_{b}} = \frac{1}{N^m_b}\sum\limits_{({\bm x}, t) \in \partial \Omega } {\left( {\left\| {{\bm u} - \frac{{\partial \bm{\delta} }}{{\partial t}} - {\bm r} \times \frac{{\partial \alpha }}{{\partial t}}} \right\|_2^2} \right)},
\\
&&{{\cal L}_i} = \frac{1}{{{N_i}}}\sum\limits_{({\bm x}, t) \in {\Gamma _i}} {\left( {\left\| {{\bm u} - {{\bm u}_i}(t)} \right\|_2^2 + \left\| {p - {p_i}(t)} \right\|_2^2} \right)},
\\
&&{{\cal L}_{data}} = \frac{1}{{N_{data}}}\sum\limits_{n = 1}^{N_{data}} {\left( {{{\left| {{\bm u}({\bm x}_u^n, t_u^n) - {\bm u}_{data}^n} \right|}^2}} + {{\left| {p({\bm x}_p^n, t_p^n) - p_{data}^n} \right|}^2} \right)},
\end{eqnarray}
where $N_f$, $N^s_{b}$, $N^m_{b}$, $N_i$ and $N_{data}$ denote the number of training points corresponding to each loss term. $\Omega$, $\Gamma _{D}$/$\Gamma _{N}$, $\partial \Omega$ and $\Gamma _i$ indicate the training domain, the stationary boundaries, the moving boundaries and the spatial-temporal coordinates for initial condition, respectively. 
Each component of the loss function is also illustrated in Fig.~\ref{fig:FCNN_Architecture}
for a clear overview.

The numbers of hidden layers and neurons in each layer of the neural network are usually selected to meet specific needs of the problem.
After trial-and-error for the sensitivity of the results, we employ 8 hidden layers and each layer contains 40 neurons universally.
We choose the continuous and differentiable hyperbolic tangent function \textit{tanh} as the activation function $\sigma(\cdot)$~\citep{lecun2015deep}.
Typically we minimize the total loss function using the Adam optimizer~\citep{kingma2014adam} with an adaptive stochastic objective function combined with the L-BFGS optimizer~\citep{liu1989limited} based on the Quasi-Newton method. 
The Adam optimizer works with a decreasing learning rate schedule. More specifically, we train the model with a learning rate of $10^{-3}$ for the first 100,000 epochs, $5\times10^{-4}$ for the subsequent 30,000 epochs, and $10^{-4}$ for the last 30,000 epochs. The L-BFGS optimizer follows for about 50,000 epochs to further diminish the residuals.
We obtain approximate solutions when the loss function reaches a sufficiently low level. 
To evaluate the accuracy of the prediction, we select the relative $L_2$ error as a metric:
\begin{equation}
{\varepsilon _V} = \frac{{{{\left\| {V - {V^*}} \right\|}_2}}}{{{{\left\| {{V^*}} \right\|}_2}}},
\end{equation}
where $V$ represents the predicted solutions $(u,v,p)$, and $V^*$ denotes the corresponding reference solutions.

All the implementations are based on the framework of DeepXDE~\citep{lu2021deepxde}, which further employs TensorFlow~\citep{abadi2016tensorflow} in the backend.
High-fidelity DNS data computed by OpenFOAM (FVM-based)~\citep{jasak2009openfoam} are taken as reference in Section~\ref{sec:Functioning as a DNS solver} and also leveraged as partial data for flow reconstruction in Section~\ref{sec:Reconstructing the flow field}. The mesh designed for each case is presented in~\ref{Appendix A}.

\section{PINNs as a direct numerical simulation solver}
\label{sec:Functioning as a DNS solver}

In this section, we employ PINNs only with initial data and validate the proposed extension as direct numerical simulation for three challenging flow problems, which involve moving boundaries. 
\subsection{In-line oscillating cylinder}
\label{subsec:In-line oscillating cylinder in fluid}
\begin{figure}[htb]
    \centering
    \includegraphics[width=1.0\linewidth]{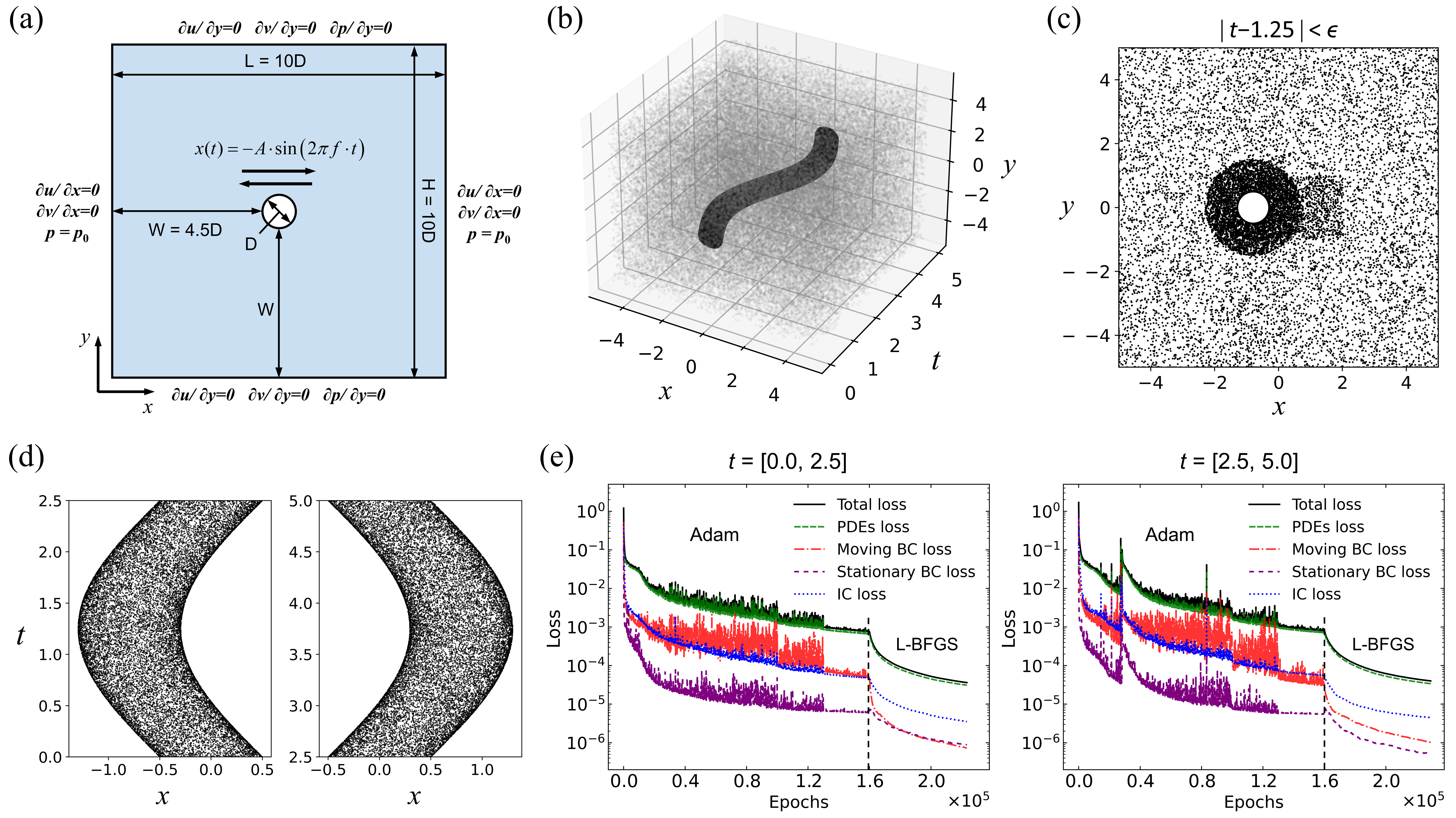}
    \caption{Problem setup and training strategy for an in-line oscillating cylinder in fluid at rest. (a) The geometry of the computational domain and boundary conditions. (b) Training points at the moving interface, around the oscillating cylinder boundary and within the computational domain in space and time. (c) Snapshots of spatial training points sampled within the temporal neighborhood $\epsilon=0.1$ at $t$=1.25. The points are sampled with finer resolution in the region close to the cylinder and within the in-line oscillation range. (d) Aerial view of the training points sampled at the moving interface in the two half-period time domains from the $y$-direction, respectively. (e) The training losses versus the number of optimization epochs in the first (left) and second (right) half periods.}
    \label{fig:inline_cyl_points_loss}
\end{figure}

The interaction of an oscillating cylinder with a fluid at rest is of particular interest in fields such as marine engineering. It is a classical case of fluid-structure interaction problem and is well-documented in literature~\citep{dutsch1998low, liu2014efficient}. The flow manifests itself by a complex vortex-structure interaction phenomena. Two key dimensionless parameters are Reynolds number $Re$ and Keulegan-Carpenter number $KC$, defined as:
\begin{equation}
Re = \frac{{{U_{max}}D}}{\nu},\quad KC = \frac{{{U_{max}}}}{{fD}},
\end{equation}
where $U_{max}$ is the maximum velocity of the moving cylinder, $D$ is the cylinder diameter, $\nu$ is the kinematic viscosity of the fluid, and $f$ is the characteristic frequency of the oscillation. The geometry of this flow is shown in Fig.~\ref{fig:inline_cyl_points_loss}(a). The translational motion of the cylinder is prescribed by a simple harmonic oscillation:
\begin{equation}
x(t) = - A \cdot \sin (2\pi f \cdot t),
\end{equation}
where $x$ denotes the location of the cylinder's center, and $A$ is the amplitude of the oscillation. Thus, the Keulegan-Carpenter number can also be written as:
\begin{equation}
KC = \frac{2 \pi A}{D}.
\end{equation}

Referring to the experiments of D{\"u}tsch et al.~\citep{dutsch1998low} and the numerical simulations of Liu et al.~\citep{liu2014efficient}, the Reynolds and Keulegan-Carpenter numbers are set to $Re=125$ and $KC=5$, respectively. Accordingly, $\rho=1$, $\nu=0.008$, $D=1$, ${U_{max}}=1$ and oscillation period $T=1/f=5$. The far-field velocities are subjected to Neumann velocity conditions,
as illustrated in Fig.~\ref{fig:inline_cyl_points_loss}(a). 
Furthermore, Dirichlet pressure condition is applied on the left and right sides,
while Neumann pressure condition is imposed on the top and bottom sides.
To capture the cylinder oscillation accurately, we enforce the Dirichlet velocity condition on the cylinder boundary, wherein $u(t)=-2\pi f \cdot A\cdot\cos(2\pi f\cdot t)$ and $v(t)=0$. As for the initial conditions, we adopt the high-fidelity velocity and pressure data obtained from the FVM at $t'=15$ (phase 0°) as the starting point $t=0$ of PINNs.

We first define the spatial computational domain as a rectangular region $[x_{min}, x_{max}] \times [y_{min}, y_{max}] = [-5, 5] \times [-5, 5]$. For the time domain, a single oscillation period from $0.0$ to $5.0$ is partitioned into two separate half-periods, $[0.0, 2.5]$ and $[2.5, 5.0]$, to be solved individually. Next, we excavate a tunnel formed by the oscillation of the cylinder in the spatial-temporal domain. A visual representation of the tunnel corresponding to one oscillation period is shown in Fig.~\ref{fig:inline_cyl_points_loss}(b). Note that the computational domain excludes the interior of the moving cylinder. Therefore, only training points obtained from locations in the domain other than the interior of the tunnel are considered. A number of $20,120$ data points from the FVM containing the velocity and pressure data of the flow field are used as the initial conditions. We randomly sample $100,000$ training points for the fluid, $20,000$ training points at the cylinder, and $10,000$ training points at the truncated rectangular boundary. We further enhance the sampling density by adding an additional $50,000$ points of finer resolution in close proximity to the surface of the cylinder, as well as $20,000$ finer points within a rectangular region $[-2, -1] \times [2, 1]$. Here, a snapshot of the distribution of the points sampled within the temporal neighborhood $\epsilon=0.1$ at $t$=1.25 is shown in Fig.~\ref{fig:inline_cyl_points_loss}(c).
In Fig.~\ref{fig:inline_cyl_points_loss}(d), an aerial view of the training points that are sampled at the moving interface in two half-periods, are looked upon in the $y$ direction. 
The convergence of the total loss, PDEs loss, moving boundary loss, stationary boundary loss, and initial condition loss, are depicted in Fig.~\ref{fig:inline_cyl_points_loss}(e),
where we observe an effective reduction of the residuals by the Adam optimizer
followed by the L-BFGS optimizer.

\begin{figure}[htb]
    \centering
    \includegraphics[width=1.0\linewidth]{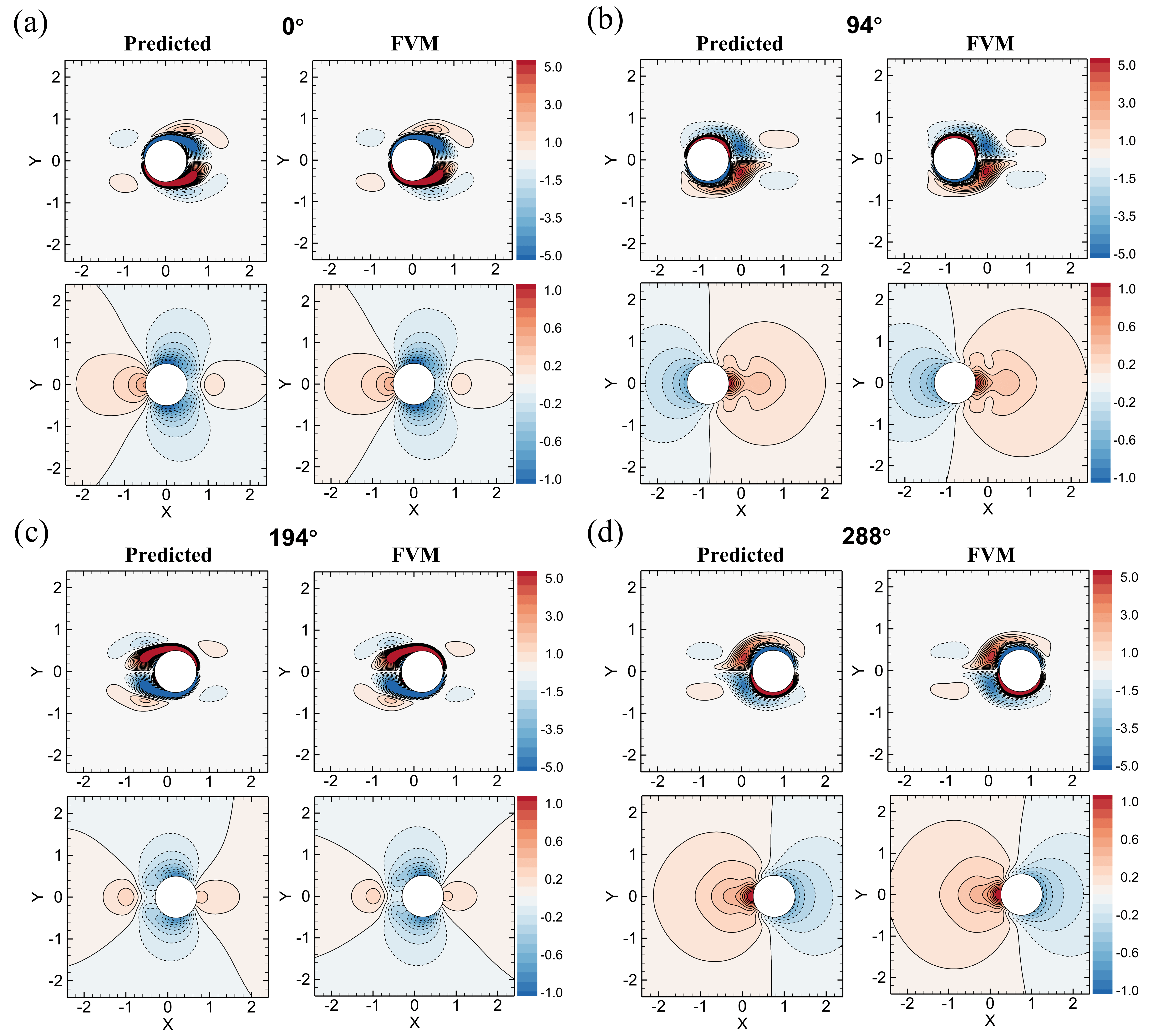}
    \caption{Results of PINNs and FVM for vorticity and pressure contours, and isolines in several typical phases. Four distinct phases and the corresponding time: (a) 0° ($t$=0); (b) 94° ($t$=1.3); (c) 194° ($t$=2.7); (d) 288° ($t$=4). Solid and dotted lines denote positive and negative contours, respectively.}
    \label{fig:inline_cyl_vorticity_p}
\end{figure}
\FloatBarrier

\begin{figure}[htb]
    \centering
    \includegraphics[width=1.0\linewidth]{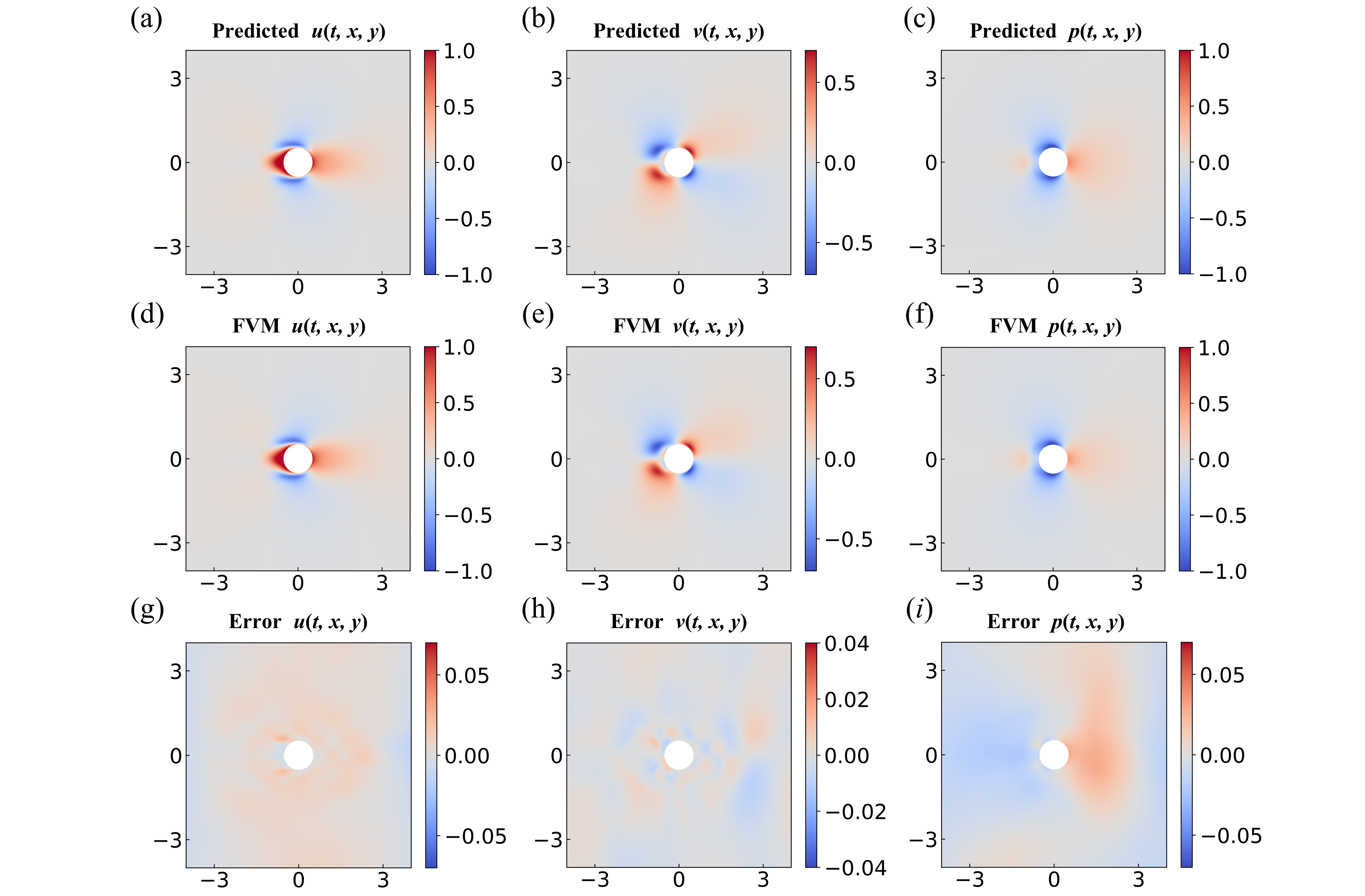}
    \caption{The comparison between PINNs and FVM for the velocity $u$, $v$ and pressure $p$ at phase 180°, where the velocity of the cylinder is maximum. The point-wise absolute errors are depicted in the bottom panels.}
    \label{fig:inline_cyl_uvp}
\end{figure}

\begin{figure}[htb!]
    \centering
    \includegraphics[width=1.0\linewidth]{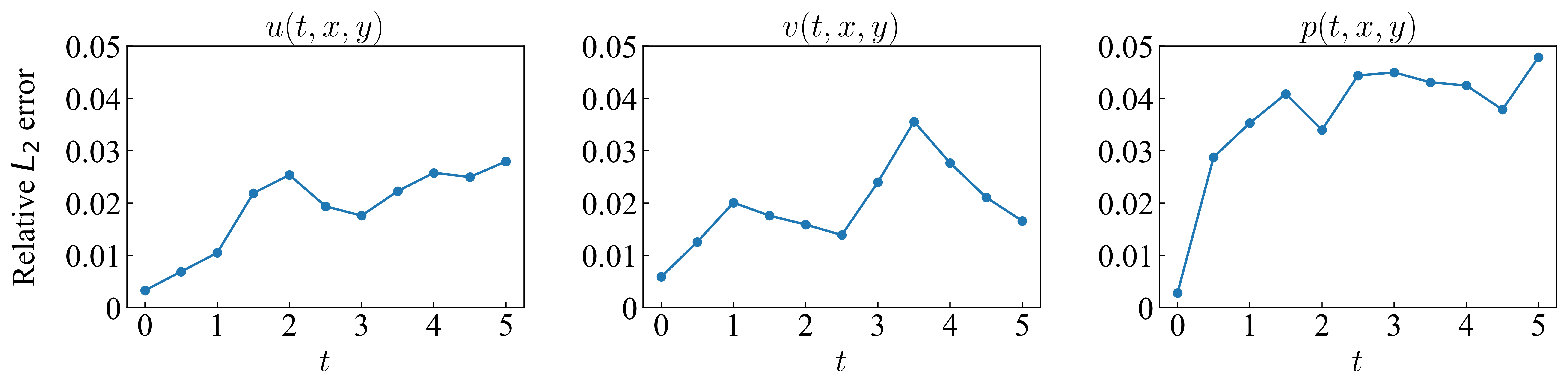}
    \caption{Relative $L_2$ errors of PINNs for the in-line oscillating cylinder in fluid at rest: the predicted velocity $u$, $v$ and pressure $p$.}
    \label{fig:inline_cyl_error}
\end{figure}

In Fig.~\ref{fig:inline_cyl_vorticity_p}, the predicted vorticity and pressure contours and isolines of the flow at four typical phases (0°, 96°, 192°, and 288°) by PINNs are presented, exhibiting excellent consistency with the results of FVM.
This indicates that the method is capable of capturing the symmetrical vortex shedding characteristics and pressure distribution during the oscillation of the cylinder. 
The first row of Fig.~\ref{fig:inline_cyl_uvp} shows the velocity $u$, $v$, and pressure $p$ at phase 180°
predicted by PINNs, where the maximum velocity of the circular cylinder is located.
These results are compared to those of FVM in the second row, and meanwhile point-wise absolute errors are presented in the third row of Fig.~\ref{fig:inline_cyl_uvp},
where the errors are well below $5\%$ overall. 
For selected time instants, we present relative $L_2$ errors for the whole field in Fig.~\ref{fig:inline_cyl_error}.
We note that the grid nodes in the FVM are employed as the coordinate points for PINNs' predictions and validations.

\begin{figure}[htb]
    \centering
    \includegraphics[width=1.0\linewidth]{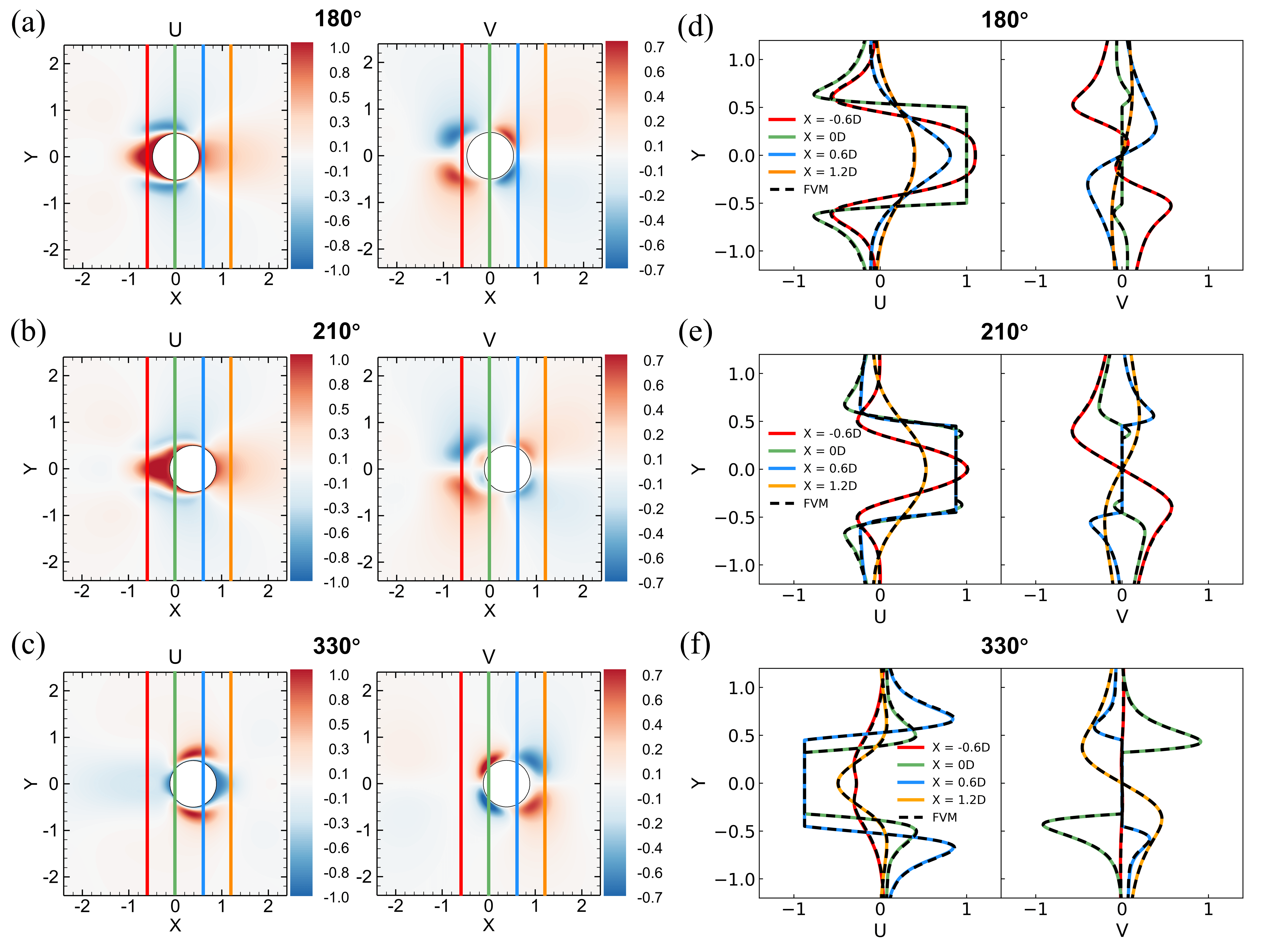}
    \caption{PINNs results for velocity contours and selected profiles for $u$ and $v$ along the $y$-direction are compared with those of FVM at three typical phases of the in-line moving cylinder. Solid and dotted lines denote PINNs and FVM results along four profiles indicated with colors: red: $x=-0.6D$; green: $x=0D$; blue: $x=0.6D$; orange: $x=1.2D$.}
    \label{fig:inline_cyl_uv_profiles}
\end{figure}
The velocity distributions in the $y$-directional cross-section are further shown in Fig.~\ref{fig:inline_cyl_uv_profiles} for $u$ and $v$ in three phases (180°, 210°, and 330°) of the oscillating cylinder, and compared with FVM results. The comparison demonstrates that the proposed model captures the finer flow distribution with great accuracy.

\subsection{Multiple cylinders translating along a circle}
\label{subsec:Multiple cylinders translating along a circle}
\begin{figure}[htb]
    \centering
    \includegraphics[width=1.0\linewidth]{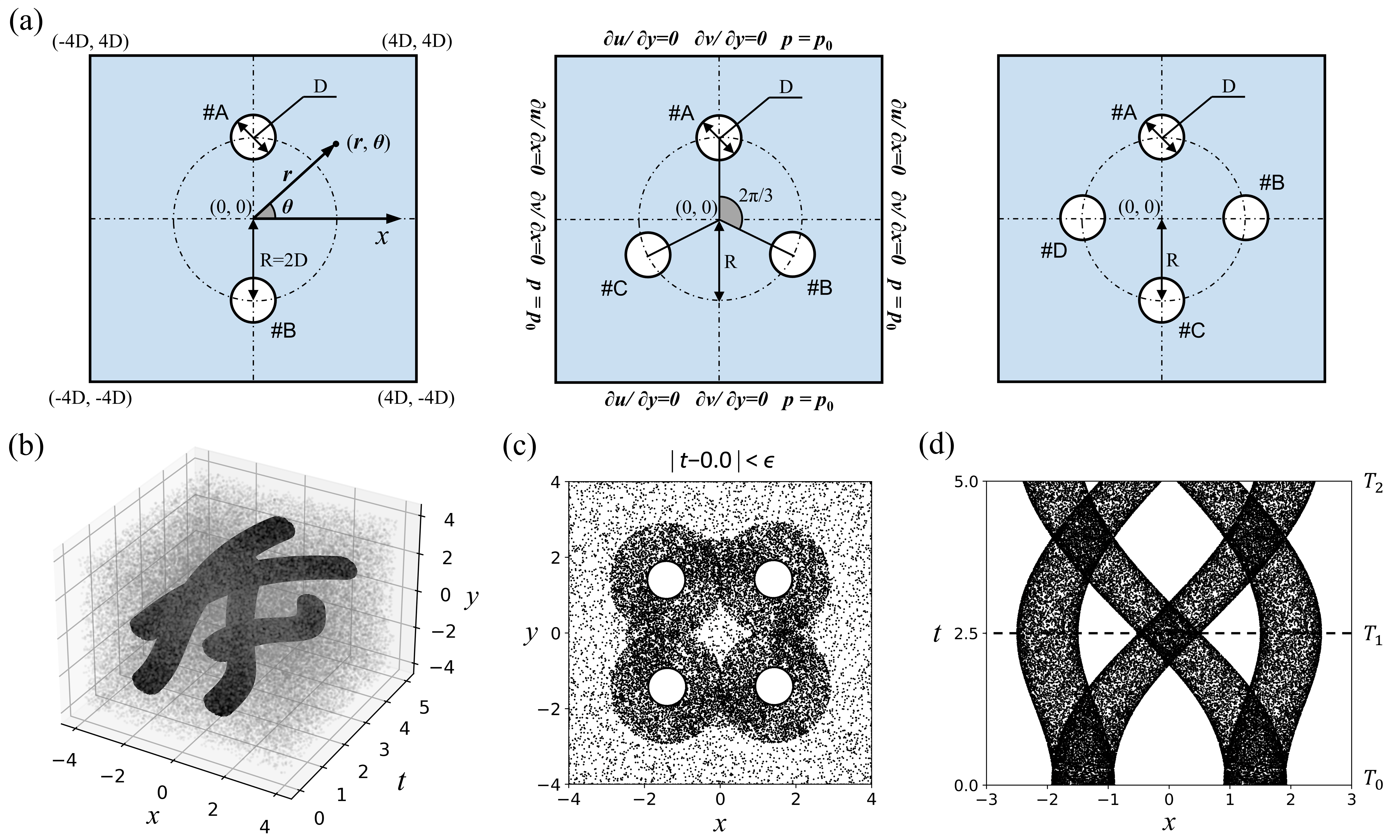}
    \caption{Problem setup and training points for multiple cylinders translating along a circle. (a) The geometry of the computational domain and boundary conditions. The left, middle, and right panels depict the geometric configurations of 2, 3, and 4 cylinders, respectively. (b) Scattered training points at four moving interfaces, around four moving boundaries, and in the fluid domain in space and time. (c) Snapshots of spatial training points sampled within a temporal neighborhood of $\epsilon=0.2$ at $t=0$. (d) Aerial view of the training points sampled at the interfaces of four moving cylinders from the $y$ direction.}
    \label{fig:multi_cyls_points}
\end{figure}

We further consider the translational motion of multiple bodies in fluid to evaluate the capability and universality of the proposed extension on PINNs. 
The key objective is to examine the relation between the number of moving boundaries and the accuracy of model training in the simulation of multiple moving bodies. Inspired by the numerical simulations of Xu et al.~\citep{xu2006immersed}, we consider 2, 3 and 4 cylinders moving clockwise in fluid along a circle of radius $R$ as an orbit. These cylinders start from rest in a sinusoidal pattern and gradually accelerate to a steady maximum velocity. The geometries of the setups, featuring 2, 3, and 4 cylinders at the maximum velocity, are presented in Fig.~\ref{fig:multi_cyls_points}(a). 

The key dimensionless parameter in this flow is the Reynolds number $Re$, defined as:
\begin{equation}
Re = \frac{{{U_{max}}D}}{\nu},
\end{equation}
where $U_{max}$ is the maximum velocity of the moving cylinders, $D$ is the diameter of the cylinders, $\nu$ is the kinematic viscosity of the fluid. The motion of the cylinders is given by:
\begin{equation}
\left\{
\begin{aligned}
    {x}& = R \cdot \cos(\theta(t)),
    \\
    {y}& = R \cdot \sin(\theta(t)),
\end{aligned}
\right.
\end{equation}
where $\theta(t)$ represents the angle of the center of the cylinder in a polar coordinate, 
which can be expressed as:
\begin{equation}
{\theta(t)} = \left\{ 
\begin{aligned}
 &\varphi + \frac{A}{R}  \cdot  \cos(2\pi f \cdot t),\quad 0 \le t \le 2.5,\\
 &\varphi - \frac{{{U_{\max }}}}{R}  \cdot  (t - 2.5),\quad t \ge 2.5,
\end{aligned}
\right.
\end{equation}
where $A$ and $f$ represent the amplitude and frequency, respectively, of the cosine acceleration, $R$ is the radius of the circle along which the cylinders are translating. The variable $\varphi$ represents the phase angle, given in polar coordinates, at which the cylinders reach their maximum linear velocity. For the two-cylinder system, the phase angles of cylinders A and B are $\varphi_A=\pi/2$ and $\varphi_B=-\pi/2$, respectively; For the three-cylinder system, the phase angles of cylinders A, B and C are $\varphi_A=\pi/2$, $\varphi_B=-\pi/6$ and $\varphi_C=-5\pi/6$, respectively; For the four-cylinder system, the phase angles of cylinders A, B, C and D are $\varphi_A=\pi/2$, $\varphi_B=0$, $\varphi_C=-\pi/2$ and $\varphi_D=-\pi$, respectively.

In all three systems, the Reynolds number is set to $Re=100$. The parameters are set as follows, $\rho=1$, $\nu=0.01$, $D=1$, $R=2$, ${U_{max}}=1$ and acceleration parameters $A=1.592$, $f=0.1$. The far-field boundaries are subjected to Neumann velocity conditions and constant pressure condition $p=1$ on four sides, as illustrated in Fig.~\ref{fig:multi_cyls_points}(a). To accurately describe the motion of the cylinder, Dirichlet velocity conditions are imposed on each boundary of the cylinder, where $u(t)=-R \cdot \sin(\theta(t)) \cdot \frac{d\theta(t)}{dt}$ and $v(t)=R \cdot \cos(\theta(t)) \cdot \frac{d\theta(t)}{dt}$. The initial condition $t=0$ for these systems is that both the fluid and the cylinders are stationary, i.e. $u=v=p=0$.

\begin{figure}[htb]
    \centering
    \includegraphics[width=1.0\linewidth]{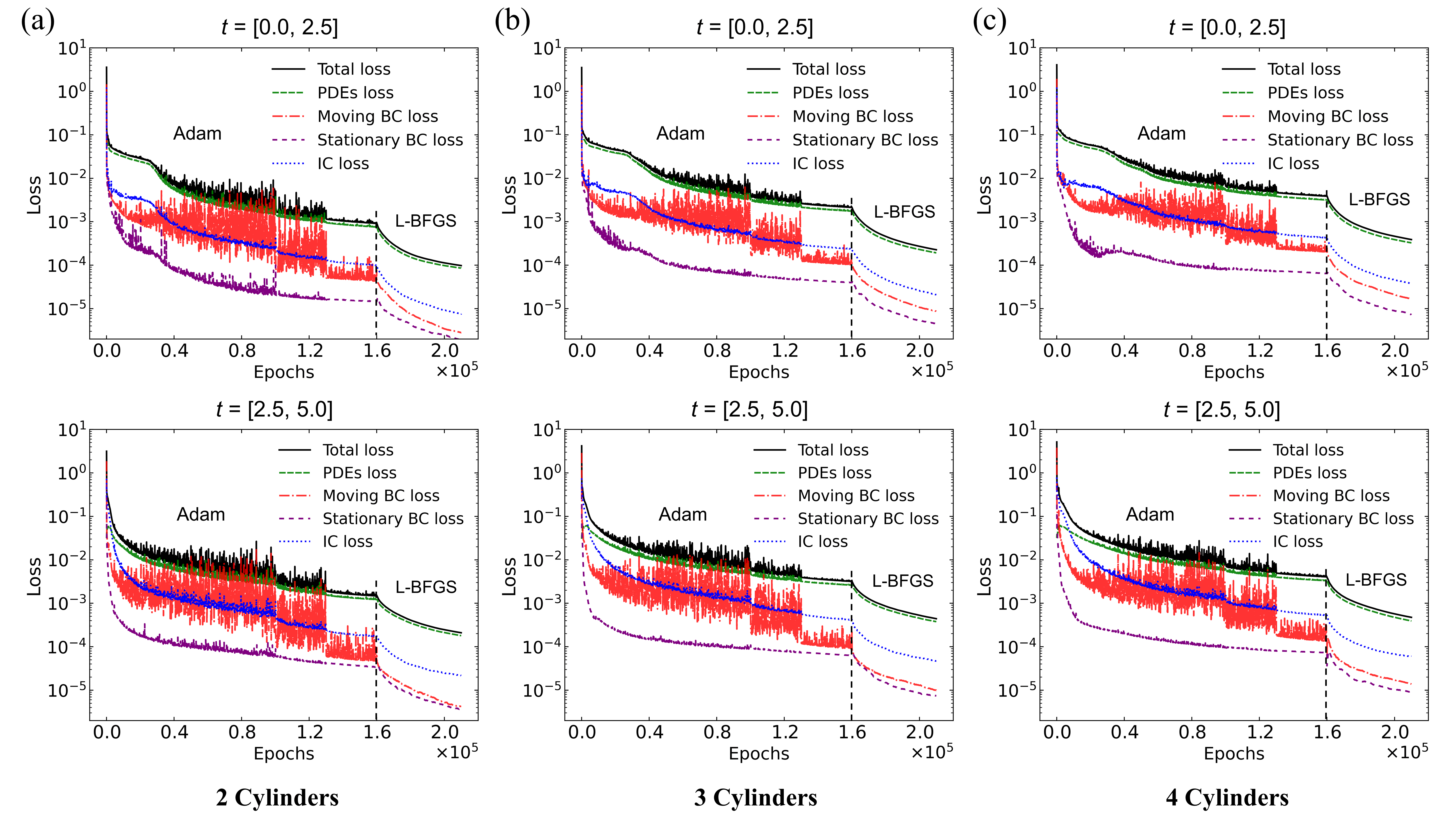}
    \caption{The training losses versus the number of optimization epochs for (a) two-cylinder system, (b) three-cylinder system, and (c) four-cylinder system. The training losses for the acceleration and steady velocity stages are depicted in the top and bottom panels, separately.}
    \label{fig:multi_cyls_loss}
\end{figure}

The spatial computational domain is defined as a rectangular region $[-4, 4] \times [-4, 4]$. The time interval [0, 5] is partitioned into two stages: [0, 2.5] for acceleration of the cylinders, and [2.5, 5] for maintaining maximum velocity through cylinder translation. And the two stages are treated as separate intervals and solved independently. Similarly, we excavate the tunnels that are formed as a result of the movements of multiple cylinders throughout the space-time domain. A visual representation of these moving boundary tunnels for the four-cylinder system is displayed in Fig.~\ref{fig:multi_cyls_points}(b). We randomly sample 70,000 points in the fluid domain, 10000 initial points, 10,000 rectangular boundary points, 15,000 points at each cylinder interface. We have enhanced the sampling density by adding an additional 40,000 points of higher resolution, in close proximity to the boundary surface of each cylinder. In addition, 50,000 points are sampled within the orbit region ($R-D/2 < r < R-D/2$) of the cylinder motion. We illustrate the spatial distribution of the training points within the temporal neighborhood where $\epsilon=0.2$ at $t$=0 for a system composed of four cylinders, as shown in Fig.~\ref{fig:multi_cyls_points}(c). Additionally, an aerial view of the training points sampled at the interfaces of four moving cylinders from the $y$ direction is depicted in Fig.~\ref{fig:multi_cyls_points}(d).

\begin{figure}[htb] 
    \centering
    \subfigure[four cylinders]{
    \includegraphics[width=0.95\textwidth]{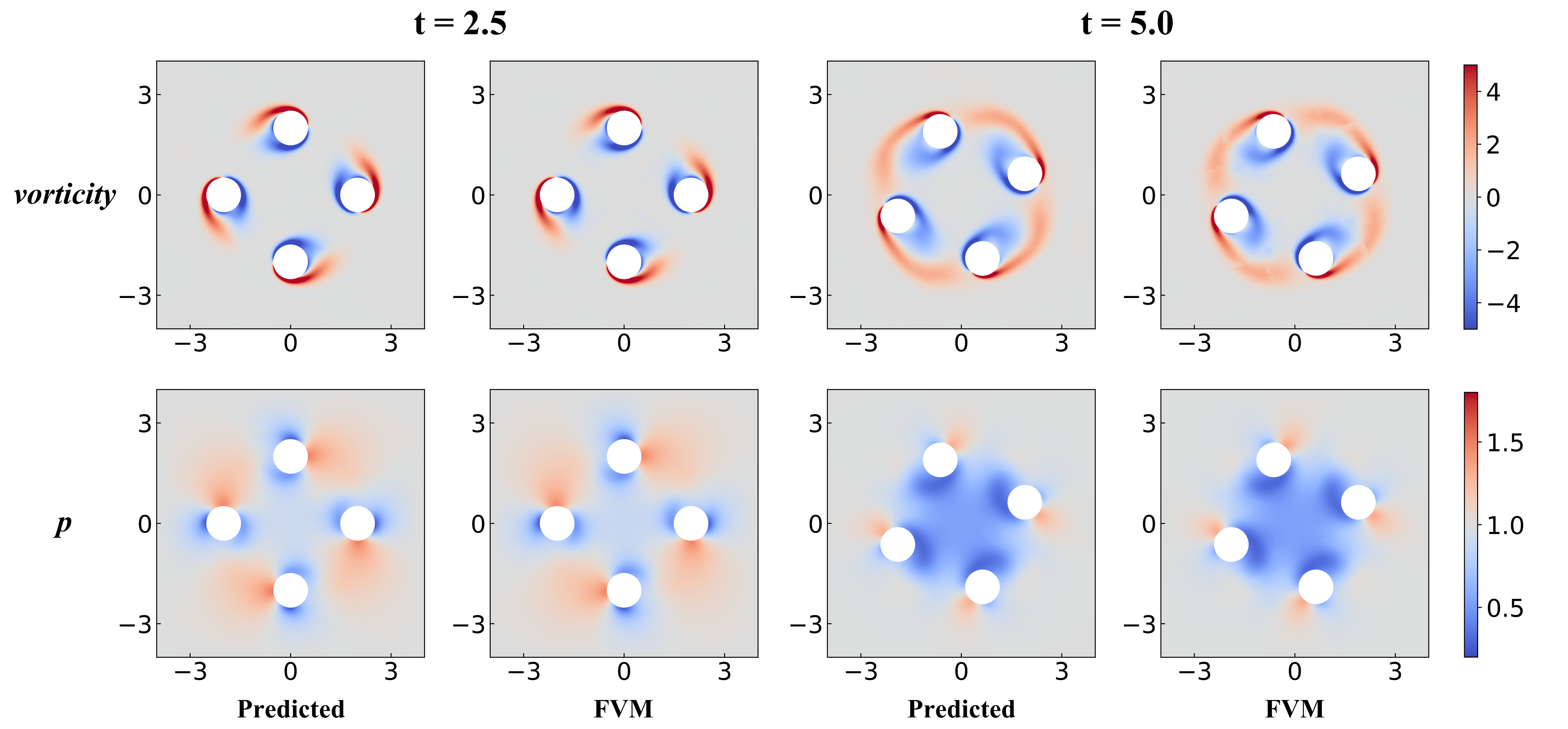}
    }
    \caption{PINNs-predicted and FVM results for vorticity and pressure contours in three multi-cylinder flow problems.}
    \label{fig:multi_cyls_vorticity_p}
\end{figure}

\begin{figure}[htb]
    \centering
    \includegraphics[width=1.0\linewidth]{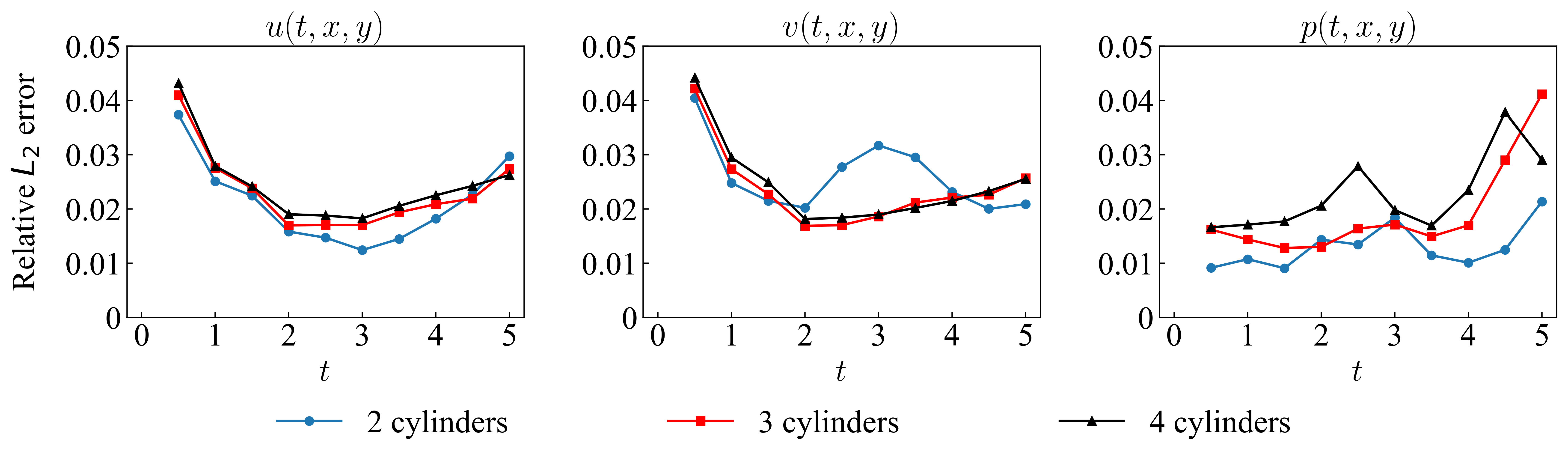}
    \caption{Relative $L_2$ errors of velocity $u$, $v$ and pressure $p$ by PINNs for multiple cylinders translating along a circle.}
    \label{fig:multi_cyls_error}
\end{figure}

Fig.~\ref{fig:multi_cyls_loss} illustrates the convergence of the total loss during training for the three multi-cylinder systems, as well as the individual components of the loss, namely PDEs loss, moving boundary loss, stationary boundary loss, and initial boundary loss.
The predicted vorticity and pressure contours of the flow for three multi-cylinder systems are presented in Fig.~\ref{fig:multi_cyls_vorticity_p}. The results demonstrate remarkable consistency with the FVM simulations, thereby confirming PINNs' effectiveness in solving multi-body flow problems. Fig.~\ref{fig:multi_cyls_error} provides the point-wise relative $L_2$ errors of predicted velocity and pressure solutions in three multi-cylinder systems. It should be noted that the grid nodes in the FVM have been selected as the coordinate points for PINNs' validations. The results from the relative errors show that the model is able to solve the flow problem containing multiple rigid bodies quite accurately. As the number of moving bodies increases, there is a slight increment of the prediction error on average. Nevertheless, the magnitude of the error remains within an acceptable range. We postulate that this error increases due to the increased complexity of the flow as a result of hydrodynamic interactions from multiple moving objects. To address this, it may become imperative to supplement PINNs with additional training points and/or explore a new network architecture to capture finer details. The primary objective of this study is to interrogate the feasibility of the new extension of PINNs to solve multi-body flow problems. Further research endeavors may potentially verify these hypotheses.
\begin{figure}[htb]
    \centering
    \subfigure[two cylinders]{
    \includegraphics[width=0.95\textwidth]{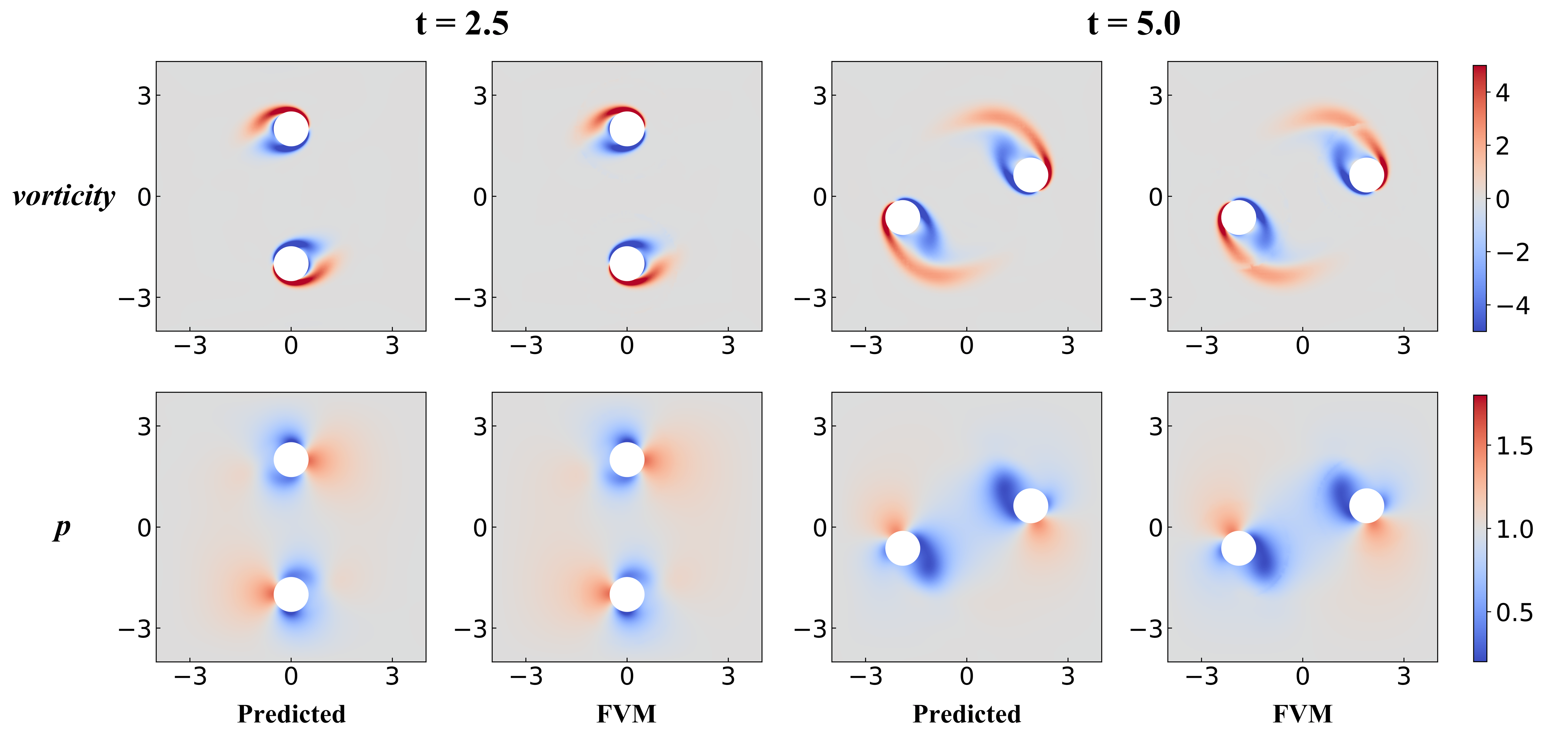}
    }
\end{figure}
\begin{figure}[htb]
    \centering
    \subfigure[three cylinders]{
    \includegraphics[width=0.95\textwidth]{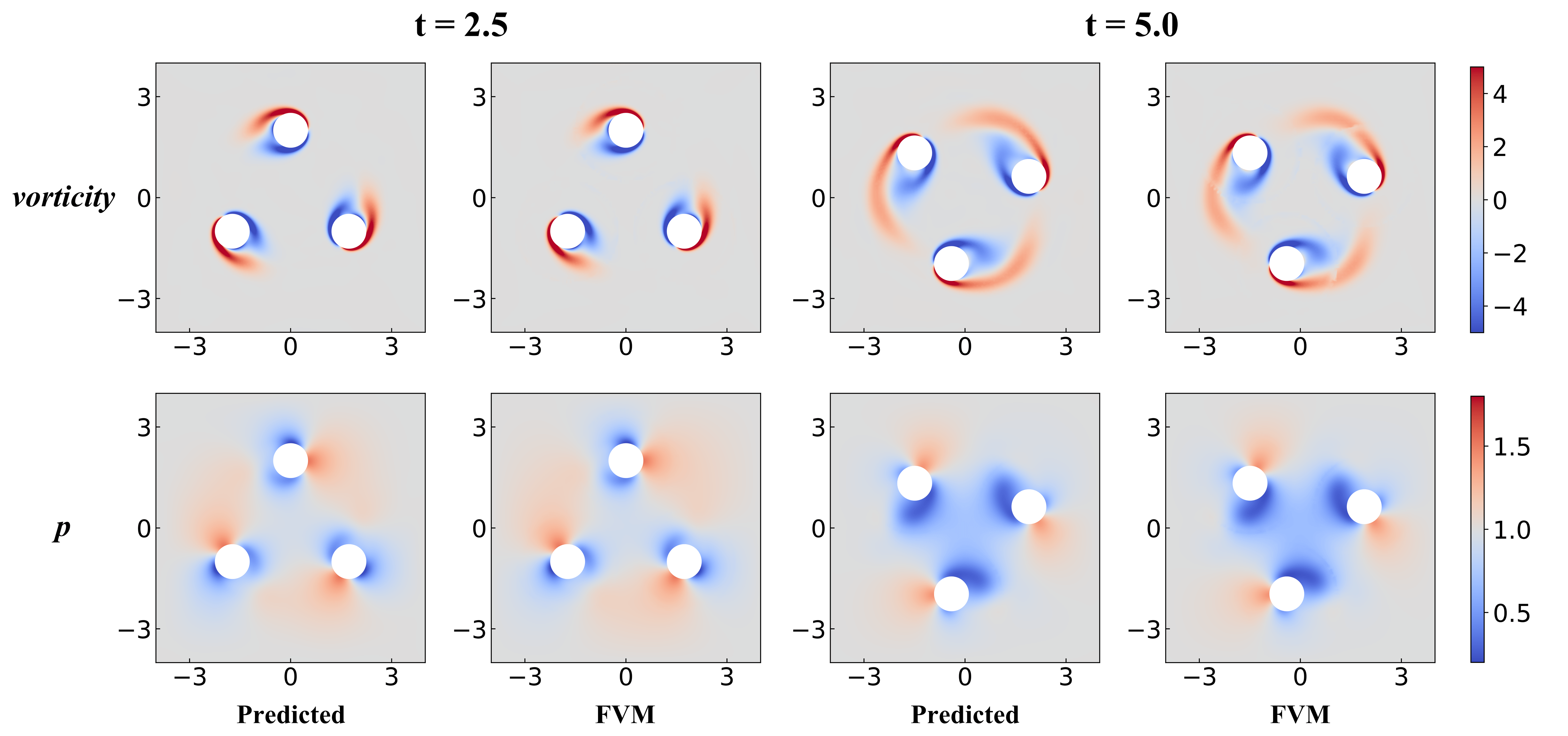}
    }
\end{figure}

\subsection{Flow around a flapping wing}
\label{subsec:DNS_Flow around a flapping wing}
\begin{figure}[htb]
    \centering
    \includegraphics[width=0.9\linewidth]{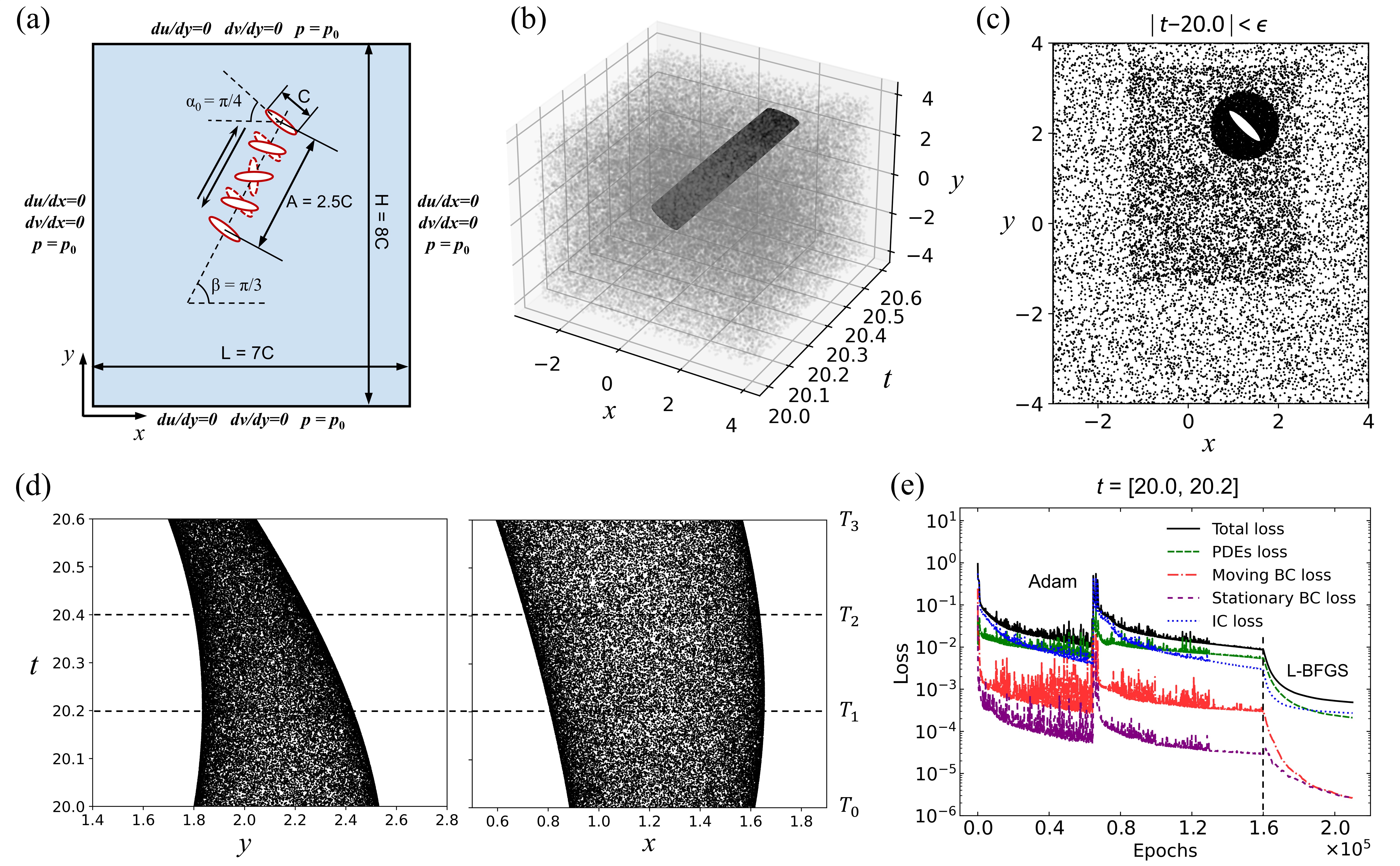}
    \caption{Problem setup and training strategy of flow around a flapping wing. (a) The geometry of the computational domain and boundary conditions. The solid ellipses indicate the down-stroke phase while the dashed ellipses represent the upstroke phase. (b) Visual representation of scattered training points at interface, around boundary and in the fluid domain. (c) Snapshots of spatial training points sampled within the temporal neighborhood $\epsilon=0.04$ at $t$=20. The training points are refined with higher resolution in the region close to the wing and within the flapping range. (d) Aerial view of all the points sampled at the wing surface from the $x$ (left) and $y$ (right) directions. (e) The training losses versus the number of optimization epochs.}
    \label{fig:flapping_wing_points_loss_DNS}
\end{figure}

We study a flapping wing that undergoes both translation and rotation. This motion pattern of the wing is similar to that of insect or bird wings during hovering. Here, we provide only the initial condition data of the flow field and to predict the entire flow field by PINNs. We consider a rigid wing with an elliptical shape, characterized by a chord length $C$ and an aspect ratio $E$. Fig.~\ref{fig:flapping_wing_points_loss_DNS}(a) depicts a sinusoidal motion of the wing cross-section in the chord direction, which is prescribed as follows:
\begin{equation}
\left\{
\begin{aligned}
    x(t)& = \frac{A}{2}\left[\cos \left(\frac{2 \pi t}{T}\right)+1\right] \cos \beta,
    \\
    y(t)& = \frac{A}{2}\left[\cos \left(\frac{2 \pi t}{T}\right)+1\right] \sin \beta,
    \\
    \alpha(t)& = \alpha_0\left[1-\sin \left(\frac{2 \pi t}{T}+\varphi\right)\right].
\end{aligned}
\right.
\end{equation}
Here $x$ and $y$ denote the coordinates of the wing's center, $\alpha$ represents the attack angle, $T$ is the flapping period, $A$ is the flapping displacement, $\beta$ is the angle of inclination of the stroke plane, and $\varphi$ is the phase angle. The velocity amplitude of the flapping wing motion is, therefore, $U_{max}= \pi \cdot A/T$. The key dimensionless parameter in this flow is the Reynolds number $Re$ defined as
\begin{equation}
Re = \frac{{{U_{max}}C}}{\nu} = \frac{{{\pi A}C}}{\nu T},
\end{equation}
where $\nu$ is the kinematic viscosity of the fluid.

The Reynolds number for is selected as $Re$=157, following the simulation undertaken by Wang et al.~\citep{wang2000two}. The various simulation parameters are set as: $C=1$, $E=4$, $A=2.5C$, $\alpha_0=\pi/4$, $\beta=\pi/3$, $\varphi=0$, $T=5$, $\rho=1$, and $\nu=0.01$. The spatial computational domain is defined as a rectangular region $[-3, 4] \times [-4, 4]$. We employ Neumann velocity conditions and a constant pressure condition $p=1$ at all far-field boundaries, as illustrated in Fig.~\ref{fig:flapping_wing_points_loss_DNS}(a). To precisely describe the wing's motion, the boundary of the wing is subject to Dirichlet velocity conditions, which can be computed using Eq.~(\ref{eq:moving boundary velocity}). As for the initial conditions, we adopt the high-fidelity velocity and pressure data obtained from the FVM at $t'=20$ as the starting point $t=20$ of PINNs,
and we solve the flow within the time domain T = [20.0, 20.6]. We use the time sequence scheme with a subdomain size of $\Delta T=0.2$. Fig.~\ref{fig:flapping_wing_points_loss_DNS}(b) displays a visual representation of the tunnel of a wing in space and time. 100,000 training points are randomly sampled in the domain composed of a rectangular region of width 7$C$ and height 8$C$ and the time subdomain. 30,000 and 20,000 points are sampled on the wing surface and rectangular boundaries respectively. 30,000 finer sampled points are added to a rectangular region $[-1.3, 1.3] \times [2.5, 3.5]$. In addition, 80,000 finer sampled points are added to a circular region with a radius of 1.5$C$ centered on the wing. There are about 8,967 data points from FVM used as initial conditions. Fig.~\ref{fig:flapping_wing_points_loss_DNS}(c) presents a snapshot of the distribution of sampled points except for initial data points within the temporal neighborhood $\epsilon=0.04$ at $t$=20.0. We additionally present, in Fig.~\ref{fig:flapping_wing_points_loss_DNS}(d), the aerial views of the training points sampled at the wing surface, are looked upon in the $x$ and $y$ direction. 
The convergence of various training losses is depicted in Fig.~\ref{fig:flapping_wing_points_loss_DNS}(e).

\begin{figure}[htb]
    \centering
    \includegraphics[width=1.0\linewidth]{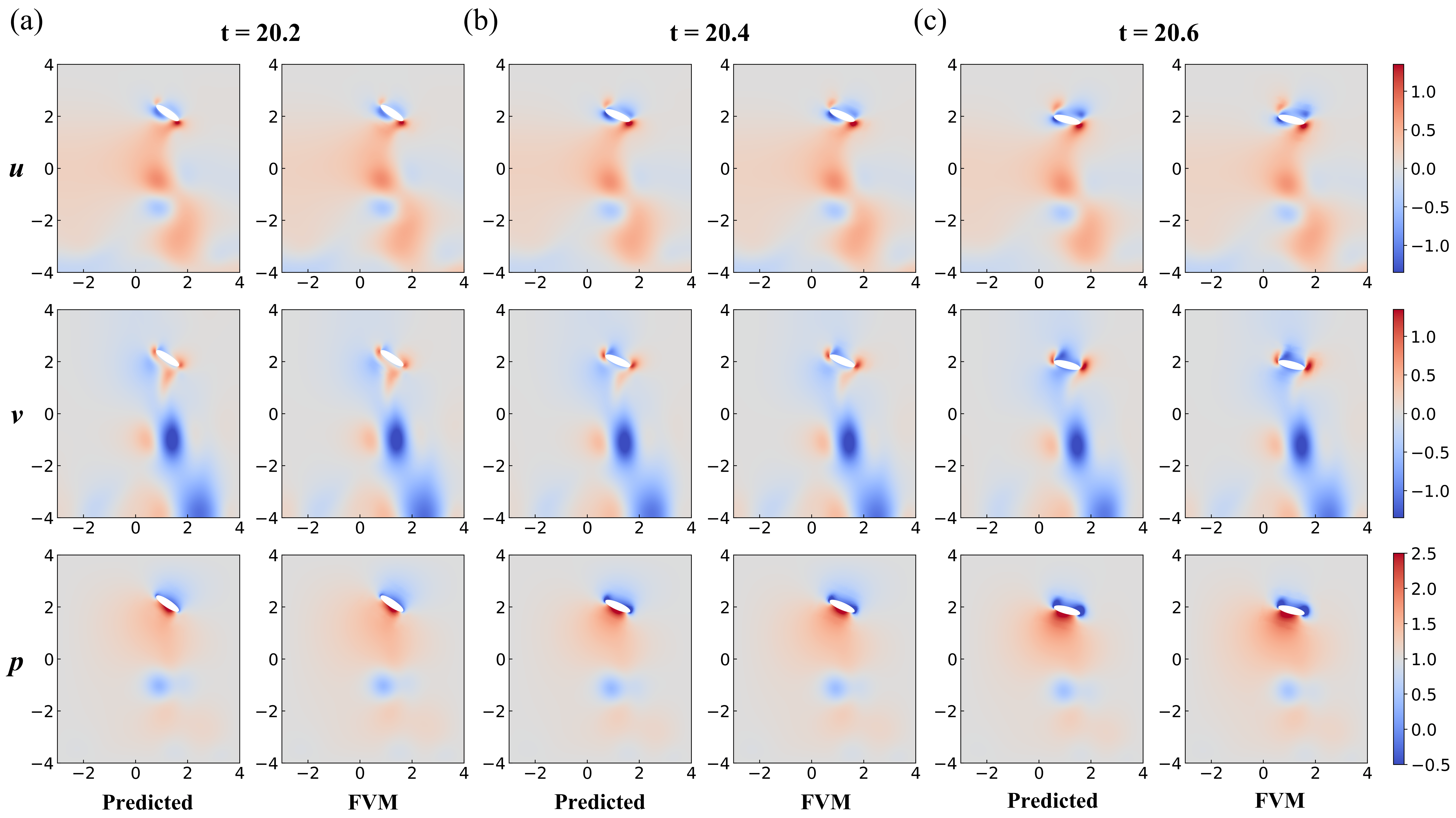}
    \caption{PINNs' results for flow around a flapping wing: velocity $u$, $v$ and pressure $p$ contours at several time frames: (a) $t$=20.2; (b) $t$=20.4; (c) $t$=20.6.}
    \label{fig:wing_uvp_DNS}
\end{figure}

\begin{figure}[htb]
    \centering
    \includegraphics[width=1.0\linewidth]{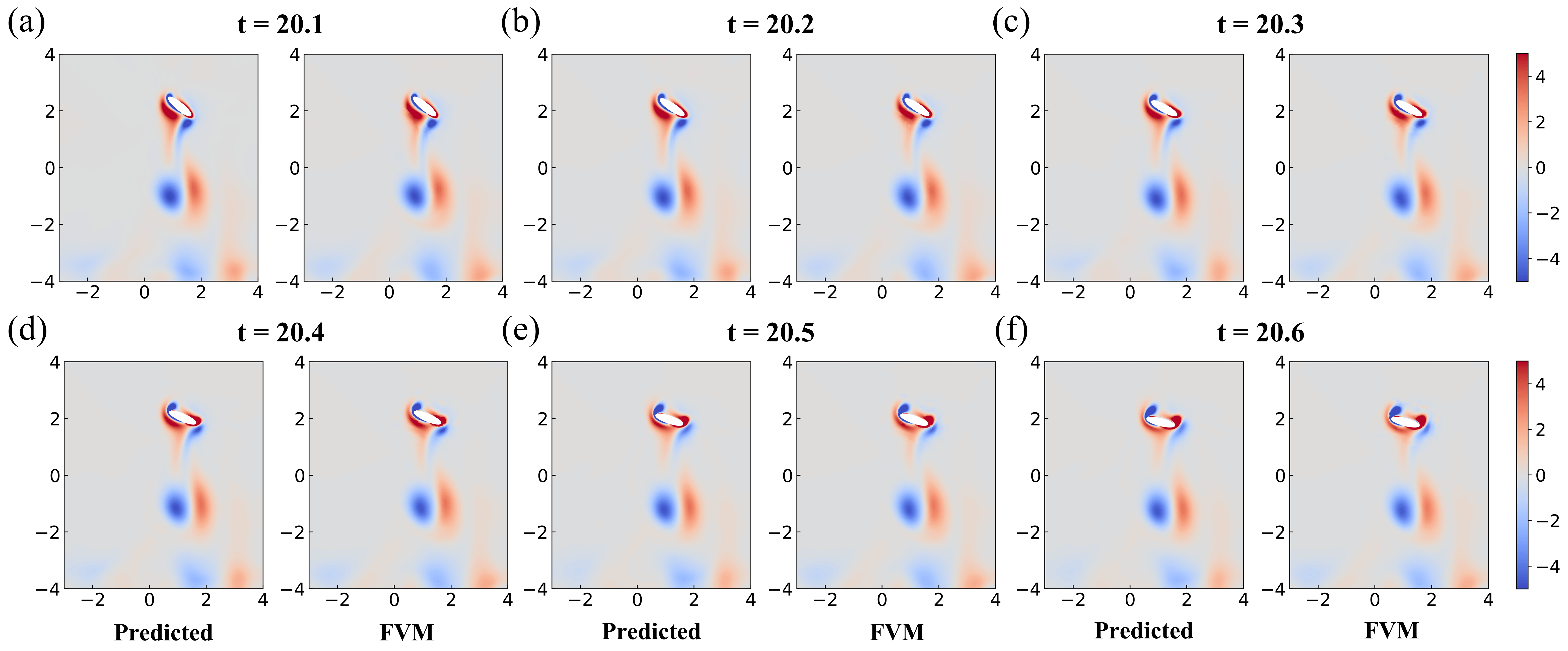}
    \caption{PINNs' results for flow around a flapping wing: vorticity contours at several time frames: (a) $t$=20.1; (b) $t$=20.2; (c) $t$=20.3; (d) $t$=20.4; (e) $t$=20.5; (f) $t$=20.6.}
    \label{fig:wing_vorticity_DNS}
\end{figure}

\begin{figure}[htb]
    \centering
    \includegraphics[width=1.0\linewidth]{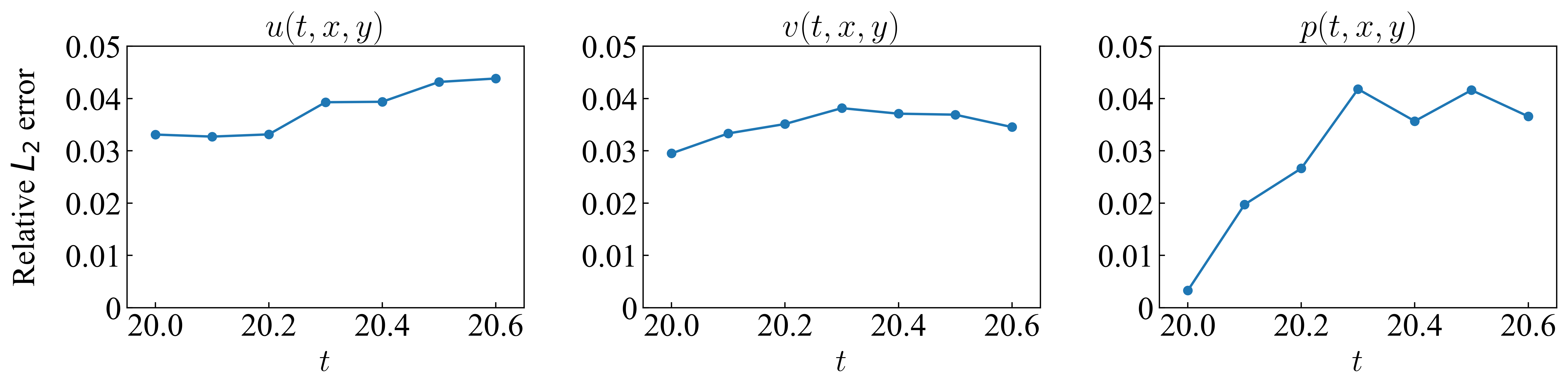}
    \caption{Relative $L_2$ error of velocity $u$, $v$ and pressure $p$ by PINNs for flow around a flapping wing.}
    \label{fig:wing_DNS_error}
\end{figure}

Fig.~\ref{fig:wing_uvp_DNS} illustrates PINNs' results of velocity $u$, $v$ and pressure $p$ contours in three time frames $t = 20.2$, $20.4$, $20.6$ within a flapping period. Fig.~\ref{fig:wing_vorticity_DNS} presents the snapshots of the vorticity contours from $t$ = 20.0 to 20.6. The model captures the typical and subtle variations in the vortex structure around the wing as it rotates and translates. Fig.~\ref{fig:wing_DNS_error} provides the point-wise relative $L_2$ errors at each time from $t$ = 20.0 to 20.6. The comparison between the predictions by PINNs and results obtained by the FVM demonstrates the capability of the former to accurately predict changes in the flow field based on the boundary conditions of the flapping wing motion. 

\section{Reconstructing the flow fields via partial data}
\label{sec:Reconstructing the flow field}
In this section, we evaluate the capability of the proposed extension in reconstructing the entire flow field when only partial data is available. 
This is also one of the best properties of PINNs,
which is unavailable in other computational frameworks.

\subsection{Transversely oscillating cylinder in steady flow}
\label{subsec:Transversely oscillating cylinder in steady flow}
\begin{figure}[htb]
    \centering
    \includegraphics[width=1.0\linewidth]{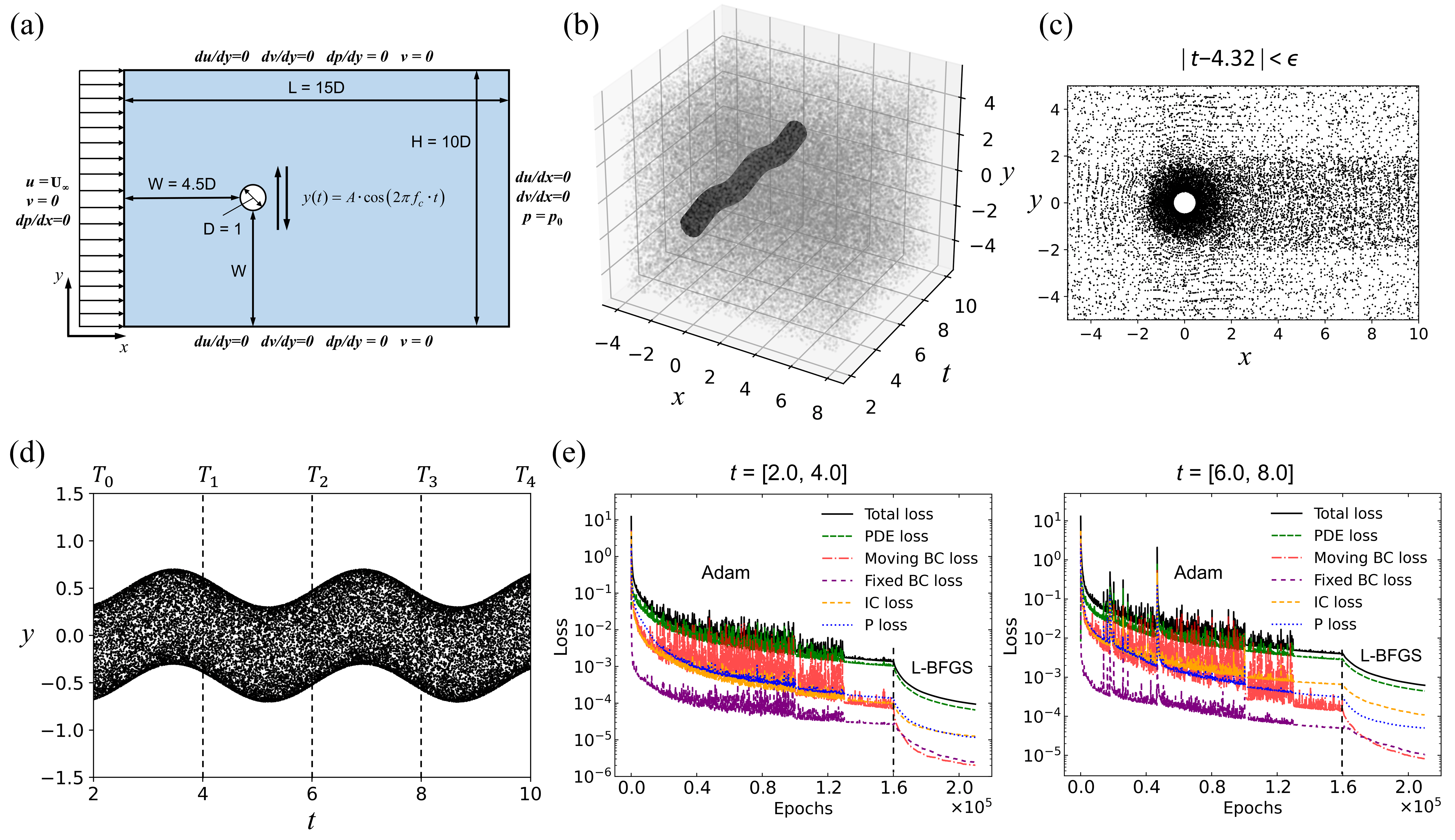}
    \caption{Problem setup and training strategy of a transversely oscillating cylinder in steady flow. (a) The geometry of the computational domain and boundary conditions. (b) Visual representation of scattered training points around the oscillating cylinder boundary and in the fluid domain in space and time. (c) Snapshots of spatial training points sampled within the temporal neighborhood $\epsilon=0.07$ at $t$=0.86. The training points are sampled with higher resolution in the vicinity of the cylinder and in the wake region where vortex shedding is prominent. (d) Aerial view of the training points sampled at the moving cylinder surface in the time domain [2, 10] from the $x$-direction. (e) The training losses versus the number of optimization epochs in two randomly selected time domains, [2, 3] and [5, 6], respectively.}
    \label{fig:trans_cyl_points_loss}
\end{figure}

We consider a transversely oscillating cylinder immersed in a steady flow. In this particular instance, the existence of asymmetrical vortex shedding introduces an elevated level of complexity in comparison to the preceding case lacking a steady flow, thus presenting an additional challenge for PINNs. The two key dimensionless parameters are Reynolds number $Re$ and Strouhal number $St$, defined as:
\begin{equation}
Re = \frac{{{U_{\infty}}D}}{\nu},\quad St = \frac{{{fD}}}{{U_{\infty}}},
\end{equation}
where $U_{\infty}$ is the velocity of uniform stream, $D$ is the diameter of the cylinder, $\nu$ is the kinematic viscosity of the fluid, and $f$ is the vortex shedding frequency. The geometry is illustrated in Fig.~\ref{fig:trans_cyl_points_loss}(a), 
where setups for inlet, outlet, and far-field boundaries are also given. 
A prescribed motion of the cylinder is expressed as:
\begin{equation}
y(t) = - A \cdot \cos (2\pi f_c \cdot t),
\end{equation}
where $y$ denotes the cross stream location of the cylinder's center, $A$ and $f_c$ are the amplitude and characteristic frequency of the oscillation, respectively.

The Reynolds number is set to $Re=185$, consistent with the numerical simulations reported by Guilmineau et al.~\citep{guilmineau2002numerical}. Accordingly, the simulation parameters are set as follows: $D=1$, the fluid density $\rho=1$, $\nu=0.01$, $U_{\infty}=1.85$, the oscillation amplitude $A/D=0.2$, and the frequency ratio $f_c/f_0=0.8$, where $f_0=St \cdot U_{\infty}/D$ is the vortex shedding frequency of the stationary cylinder at $St=0.195$. Neumann velocity and pressure boundary conditions are used at the far-field boundaries on the top and bottom sides. At the inlet boundary, a steady horizontal free stream is imposed with the velocity Dirichlet boundary condition $u=U_{\infty}$ and $v=0$. And the zero pressure condition and Neumann velocity conditions are applied to the outlet. Dirichlet velocity conditions are imposed on the cylinder boundary to capture the no-slip condition due to oscillation. Specifically, we set $u(t)=0$ and $v(t)=2\pi f_c \cdot A\cdot\sin(2\pi f_c \cdot t)$. We take the velocity and pressure data on 21,535 points obtained from the FVM at $t'=2.0$ as the initial conditions to train the model from $t=2.0$.

\begin{figure}[htb]
    \centering
    \includegraphics[width=0.8\linewidth]{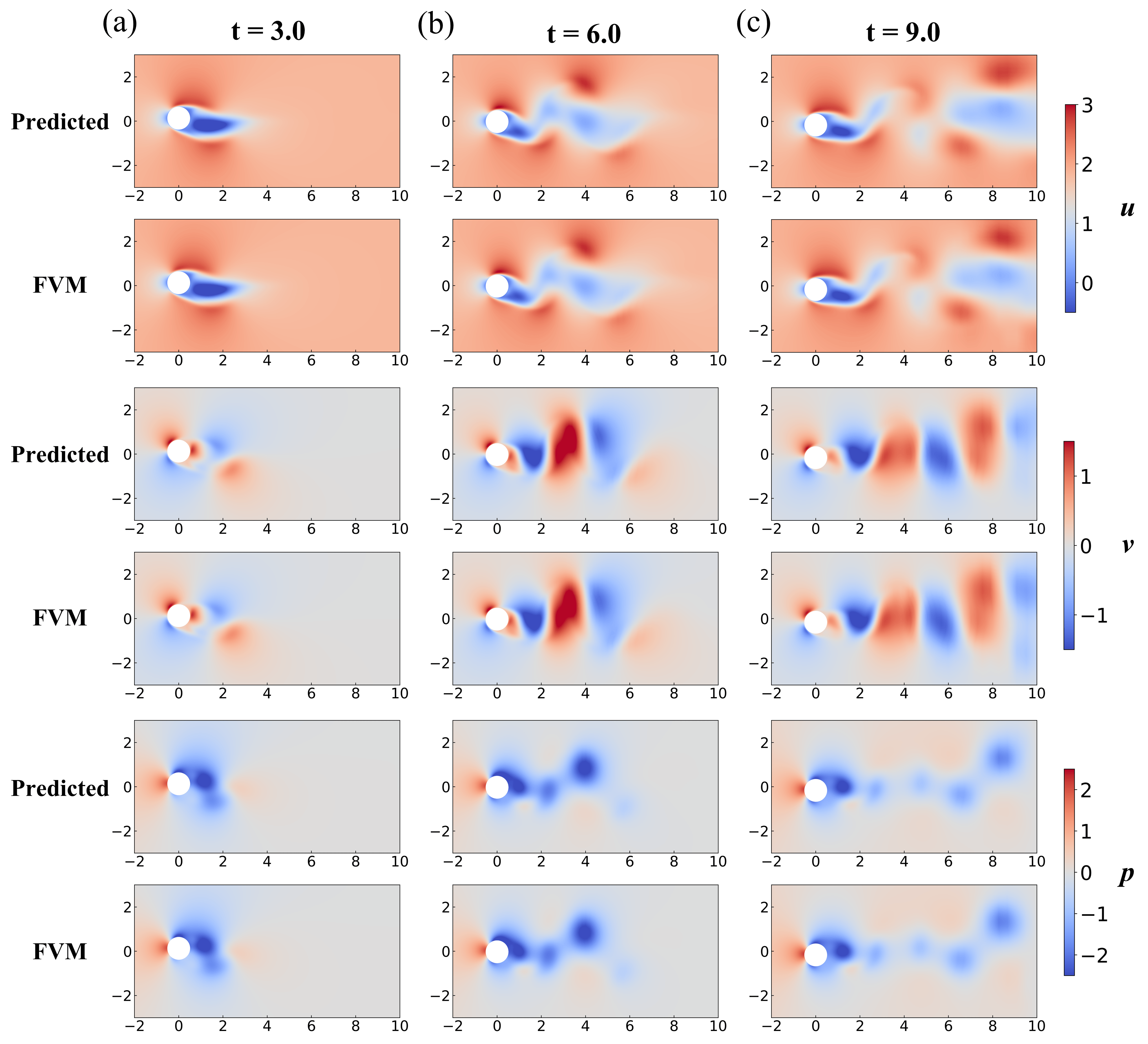}
    \caption{PINNs-predicted and FVM results for velocity $u$, $v$ and pressure $p$ contours for several time frames. (a) $t$=1.0; (b) $t$=4.0; (c) $t$=7.0.}
    \label{fig:trans_cyl_uvp}
\end{figure}

We first define the computational spatial domain, i.e., a rectangular region $[-5, 10] \times [-5, 5]$. As vortex shedding takes time to develop, we adapt a time sequence training scheme in which the time domain [2, 10] is discretized into $n$ domains as: [$T_0=2$, $T_1$], [$T_1$, $T_2$], $\cdots$, [$T_{n-1}$, $T_n=10$], to be solved individually. We set the range of the time sub-domain to $\Delta T=2$. Next, we excavate the tunnel formed by the oscillation of the cylinder in the whole spatial-temporal domain. A visual representation of the tunnel corresponding to about two oscillation periods (one period is 3.465) is shown in Fig.~\ref{fig:trans_cyl_points_loss}(b). For each sub-domain [$T_{n-1}$, $T_n$], a total of 452,235 FVM data points, corresponding to 21 time snapshots, scatter in space and time, with a time interval of $\Delta t$ = 0.1 between two consecutive snapshots. Among these FVM data points, a total of 150,000 data points are randomly selected to reconstruct the entire field.
In each sub-domain [$T_{n-1}$, $T_n=T$], we randomly sample 50,000 domain points, 10,000 cylinder boundary points, and 10,000 rectangular boundary points. Moreover, we additionally sample 30,000 finer points in the vicinity of the cylinder and 30,000 points in the wake region where vortex shedding is prominent. Fig.~\ref{fig:trans_cyl_points_loss}(c) displays a snapshot of the distribution of sampled points from the above-mentioned processes within the temporal neighborhood $\epsilon=0.07$ at $t$=4.32. Furthermore, we provide an aerial perspective of the training points that are sampled on the moving cylinder surface in $t$=[2, 10], from the $x$ direction, as shown in Fig.~\ref{fig:trans_cyl_points_loss}(d).


\begin{figure}[htb]
    \centering
    \includegraphics[width=1.0\linewidth]{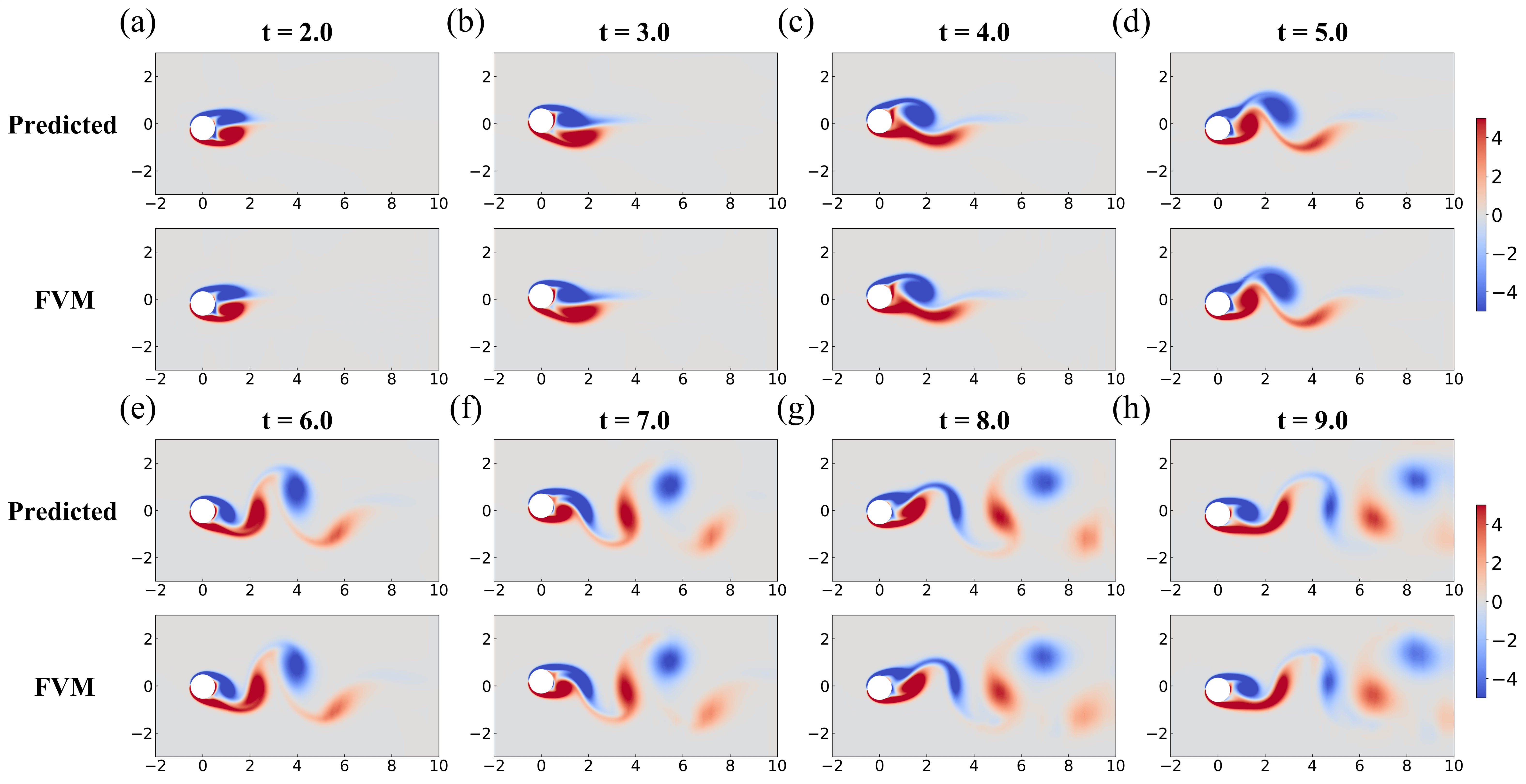}
    \caption{Results of PINNs and FVM for vorticity contours in several time frames. (a) $t$=0.1; (b) $t$=1.0; (c) $t$=2.0; (d) $t$=3.0; (e) $t$=4.0; (f) $t$=5.0; (g) $t$=6.0; (h) $t$=7.0. A significant vortex shedding phenomenon is observed. }
    \label{fig:trans_cyl_vorticity} 
\end{figure}

\begin{figure}[htb!]
    \centering
    \includegraphics[width=1.0\linewidth]{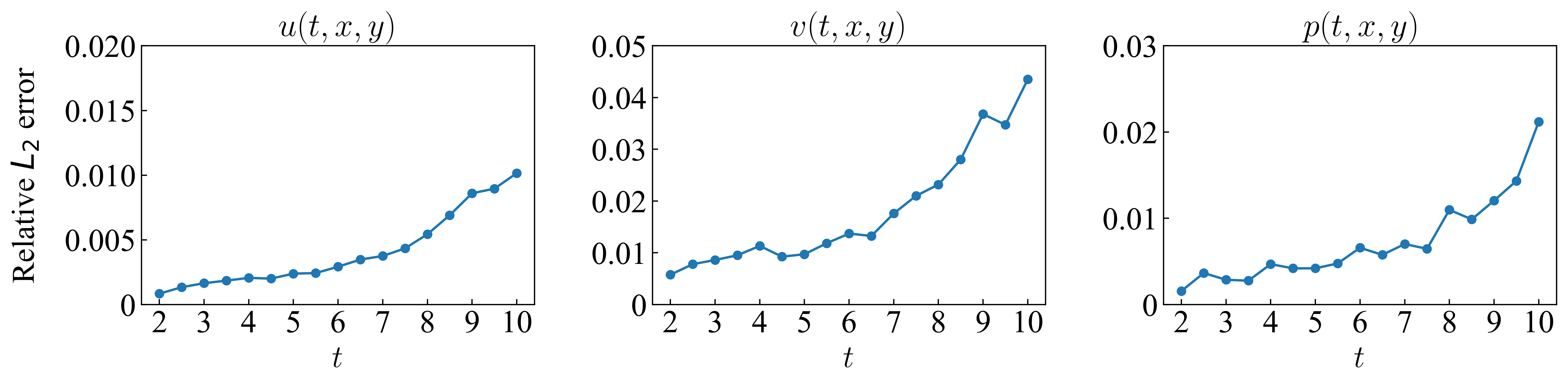}
    \caption{Relative $L_2$ errors of PINNs for the transversely oscillating cylinder in steady flow: the inferred velocity $u$, $v$ and pressure $p$.}
    \label{fig:trans_cyl_error}
\end{figure}

\begin{figure}[htb]
    \centering
    \includegraphics[width=1.0\linewidth]{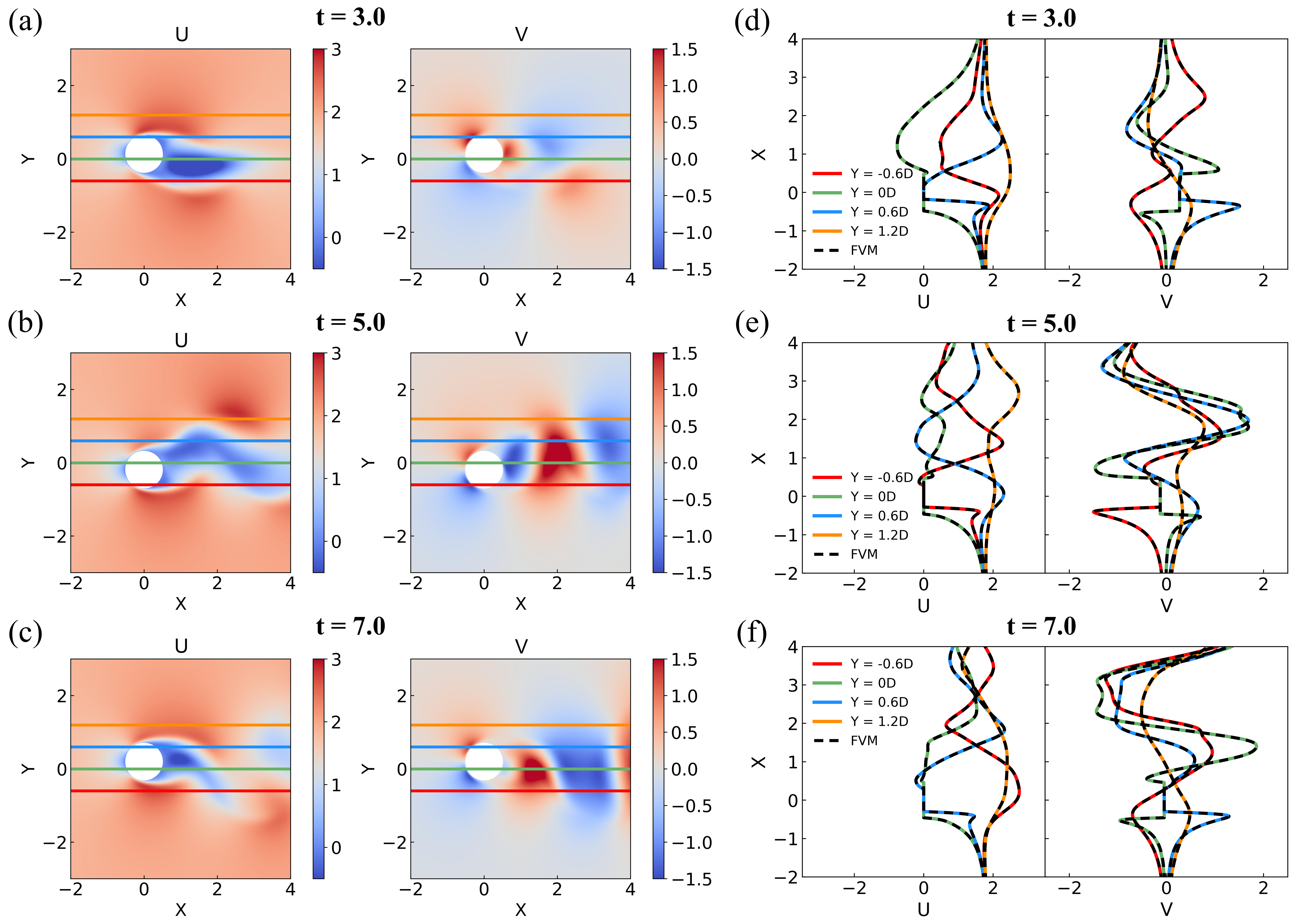}
    \caption{The PINNs-predicted velocity contours and the velocity $u$, $v$ distribution in the $y$-direction for several frames of the cylinder, and are compared with the FVM results. Solid and dotted lines denote PINNs and FVM results in four profiles(indicated with colors: red: $y=-0.6D$; green: $y=0D$; blue: $y=0.6D$; orange: $y=1.2D$).}
    \label{fig:trans_cyl_uv_profiles}
\end{figure}

Fig.~\ref{fig:trans_cyl_points_loss}(e) illustrates the convergence of the total loss and its individual components during training, including the PDEs loss, moving boundary loss, stationary boundary loss, and initial boundary loss.
Fig.~\ref{fig:trans_cyl_uvp} shows the predicted velocity and pressure at time instances $t$=1.0, 4.0, and 7.0. Additionally, Fig.~\ref{fig:trans_cyl_vorticity} illustrates the vorticity predictions of the trained model for the time period of t=[0, 7] and compared with the FVM results. The comparison of vorticity indicates that the model successfully reproduces the vortex shedding phenomenon in alignment with the FVM calculation. Despite the slightly larger dissipation in the wake region, the predicted solution accurately and comprehensively captures the process of vortex shedding from the cylinder and its development in the wake region. Fig.~\ref{fig:trans_cyl_error} provides the point-wise relative $L_2$ errors of inferred velocity and pressure solutions at each time. The relative $L_2$ errors illustrate the rapid accumulation of errors in time series model predictions over time. Offering initial conditions with high precision may mitigate errors to a certain extent in the early stages of the simulation, but these errors will surge rapidly in the subsequent stages. The velocity distributions in the $x$-directional cross-section are further shown in Fig.~\ref{fig:trans_cyl_uv_profiles} for $u$ and $v$ in three time frames ($t$ = 3.0, 5.0, and 7.0) of the oscillating cylinder, and compared with FVM results. The comparison shows that the proposed model accurately captures finer flow distributions. Overall, our strategy for PINNs is capable of reproducing the phenomenon of vortex shedding with an acceptable margin of error.

\FloatBarrier

\subsection{Flow around a flapping wing}
\label{subsec:reconst_Flow around a flapping wing}
In this case, we still consider a flapping wing that undergoes both translational and rotational motions. We solve this problem to demonstrate the flexibility of our proposed framework in the moving boundary scenario, that is, reconstruct the entire flow field (velocity $v$ and pressure $p$) based on the partial data (velocity $u$) of the spatial-temporal domain of the flow field.

\begin{figure}[htb]
    \centering
    \includegraphics[width=1.0\linewidth]{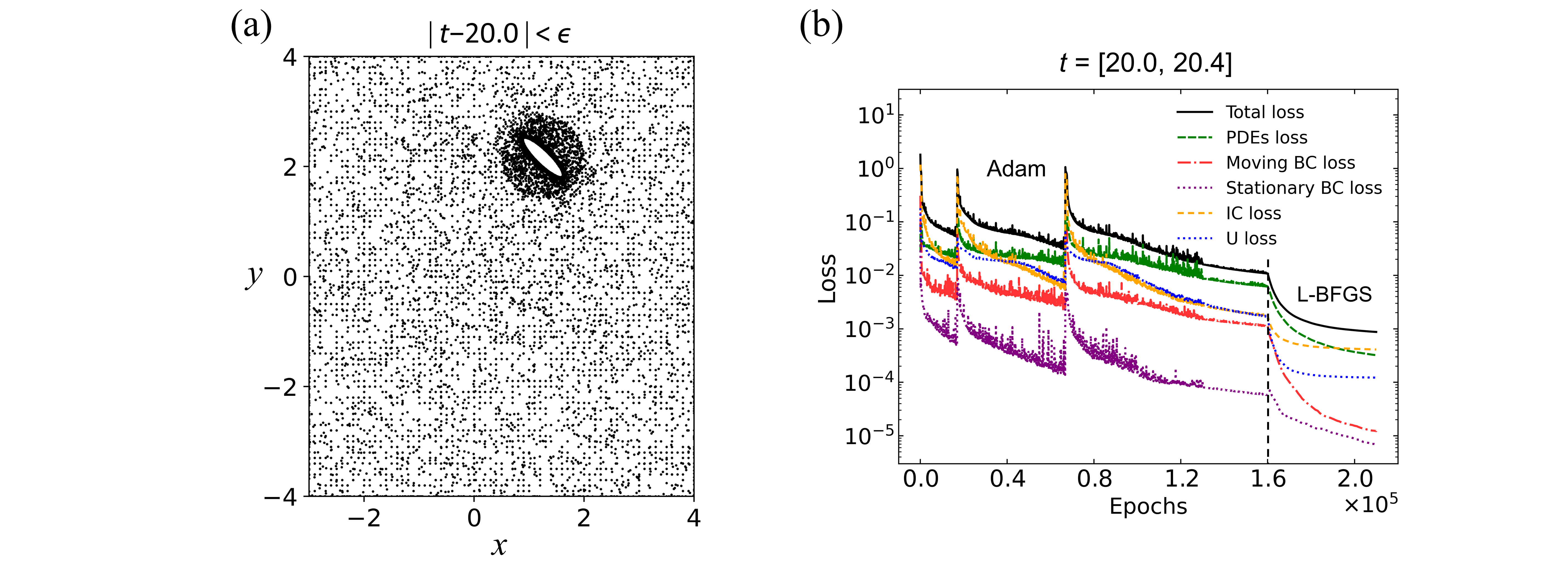}
    \caption{Training strategy for the reconstruction of the flow field around a flapping wing. (a) Snapshots of data points randomly selected and training points sampled within the temporal neighborhood $\epsilon=0.02$ at $t$=20. (b) The training losses versus the number of optimization epochs.}
    \label{fig:flapping_wing_points_loss_reconst}
\end{figure}

Here, we employ the velocity $u$ data at the grid points of the FVM as Dirichlet constraint to train the neural network. Meanwhile, we take the velocity and pressure data on $8,967$ points obtained from the FVM at $t'=20.0$ as the initial conditions to train the model from $t=20.0$. A single neural network is used to reconstruct the flow field in the time domain [20.0, 20.4]. A total of 367,738 FVM data points, corresponding to 41 time snapshots, scatter in space and time, with a time interval of $\Delta t$ = 0.01 between two consecutive snapshots. Among these FVM data points, a total of 120,000 data points are randomly selected to reconstruct the entire field. In addition, 50,000 collocation points are randomly sampled in the domain to complement the resolution of the computational domain. 25,000 and 10,000 points are sampled on the wing surface and rectangular boundaries, respectively. 40,000 finer sampled points are sampled within a circular region with a radius of 1.5$C$ centered on the wing. 10,000 finer sampled points are collected within a rectangular region $[-1.3, 1.3] \times [2.5, 3.5]$. Fig.~\ref{fig:flapping_wing_points_loss_reconst}(a) depicts a snapshot at $t$=20.0 demonstrating the distribution of data points and sampled collocation points. 

\begin{figure}[htb]
    \centering
    \includegraphics[width=0.95\linewidth]{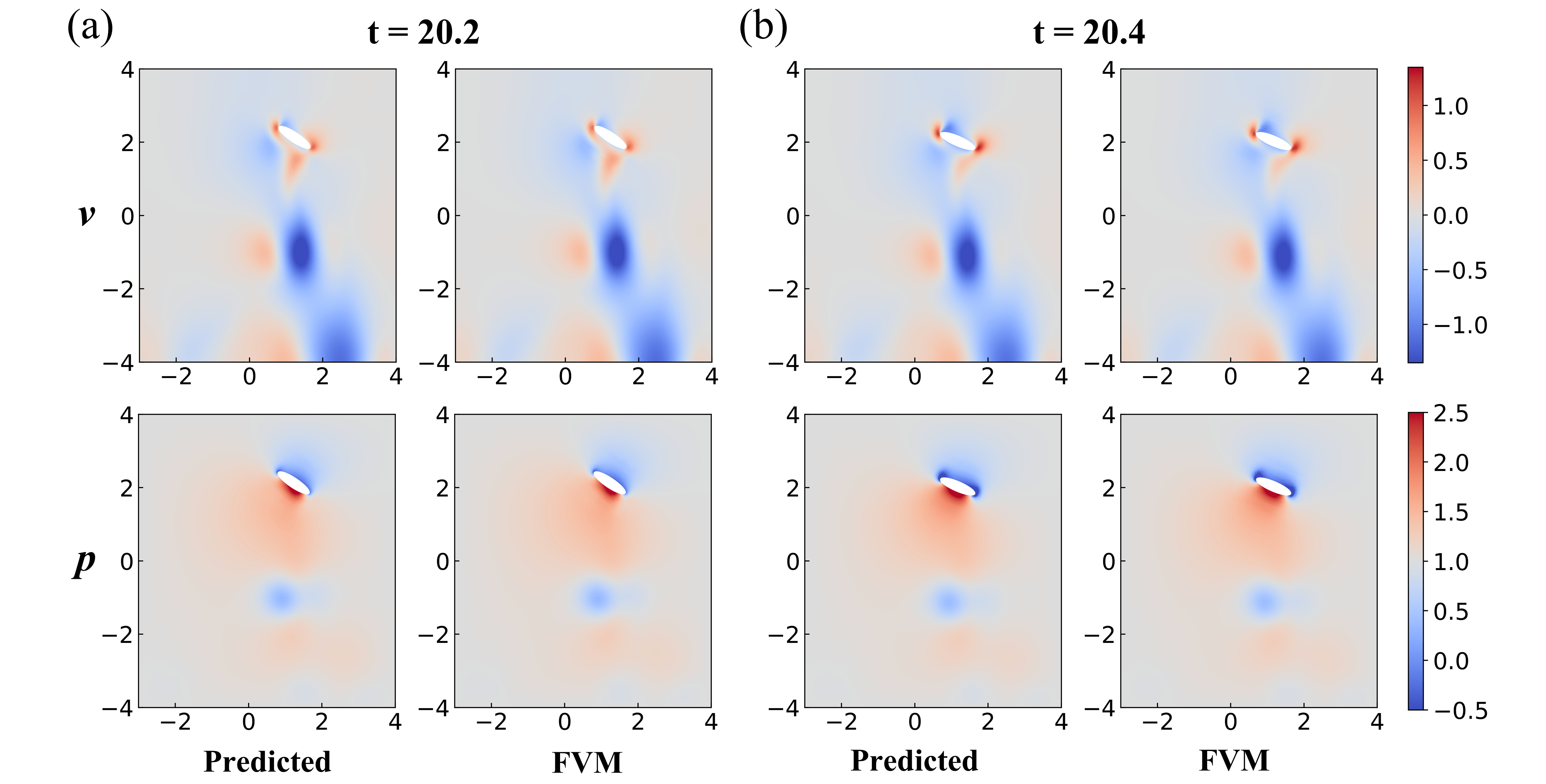}
    \caption{Flow around a flapping wing: inferred velocity $v$ and pressure $p$ contours at two time frames: (a) $t$=20.2; (b) $t$=20.4.}
    \label{fig:wing_inferred_vp_reconst}
\end{figure}

\begin{figure}[htb]
    \centering
    \includegraphics[width=1.0\linewidth]{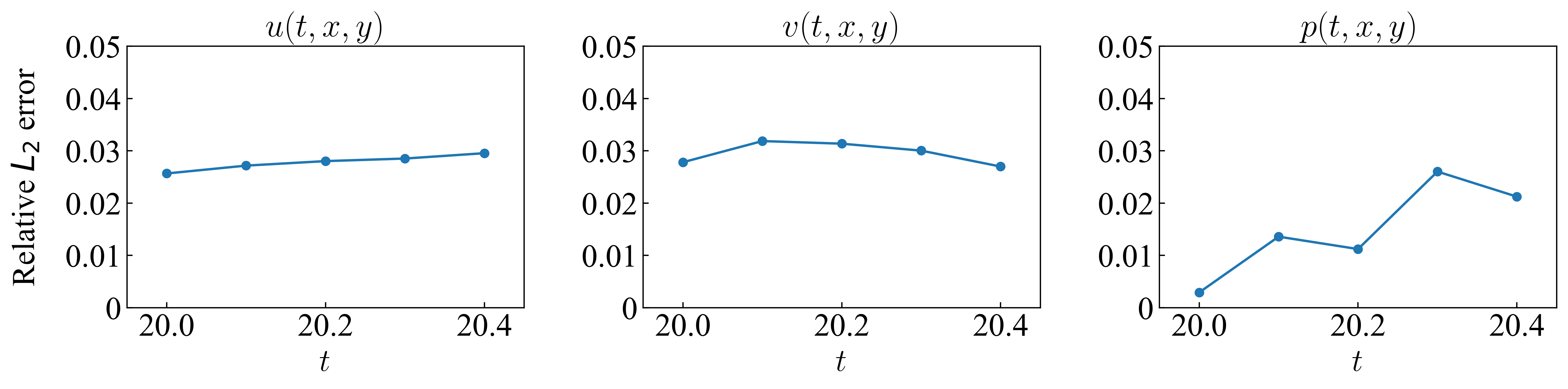}
    \caption{Relative $L_2$ errors of PINNs for flow around a flapping wing: the inferred velocity $u$, $v$ and pressure $p$.}
    \label{fig:wing_reconst_error}
\end{figure}

And the convergence of various training losses is given (see Fig.~\ref{fig:flapping_wing_points_loss_reconst}(b)).
Fig.~\ref{fig:wing_inferred_vp_reconst} shows the snapshots of inferred velocity $v$ and pressure $p$ contours in two time frames $t = 20.2$, $20.4$. Fig.~\ref{fig:wing_reconst_error} provides the point-wise relative $L_2$ errors at each time from $t$ = 20.0 to 20.4. Obviously, providing partial data for the entire flow field can significantly reduce the relative error in velocity and especially pressure compared to providing data only of the initial conditions. This demonstrates the flexibility of our proposed strategy in solving moving boundary scenarios, which can rely on different types or amount of data for the prediction and reconstruction of the entire flow field.

\FloatBarrier

\subsection{A cylinder settling under gravity}
\label{subsec:A cylinder settling under gravity}
In the last case, we conduct simulations pertaining to the free settling of a cylinder under gravity. This instance can be classified as a two-way fluid-structure interaction problem. In this case, the settling trajectory of the cylinder is extracted from the FVM results, making this problem degenerate into a flow problem with prescribed motion to be solved using our proposed strategy.
\begin{figure}[htb!]
    \centering
    \includegraphics[width=1.0\linewidth]{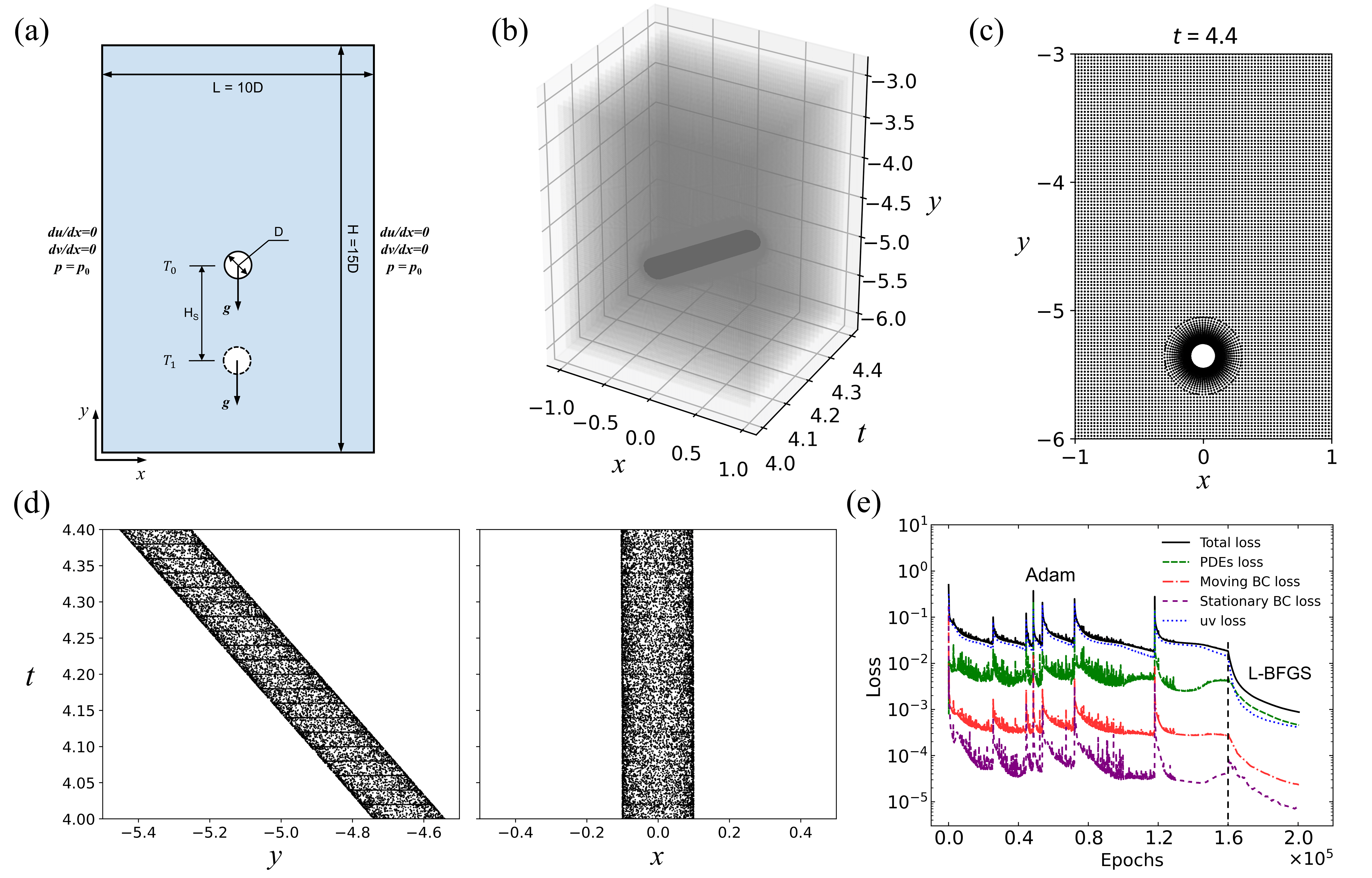}
    \caption{Problem setup and training strategy of a cylinder settling under gravity. (a) The geometry of the computational domain and boundary conditions. (b) Visual representation of scattered training points around the moving boundary and grid points in space and time. (c) Snapshots of the distribution of data points at $t=4.4$. The spatial resolution of the points is gradually enhanced in the proximity of the cylinder region. (d) Aerial view of all the points sampled at the moving cylinder surface from the $x$ (left) and $y$ (right) directions. (e) The training losses versus the number of optimization epochs.}
    \label{fig:cyl_settling_points_loss}
\end{figure}

The objective is to employ partial flow field information for the reconstruction of the complete flow field, with the pressure $p$ inferred from the known velocities $u$ and $v$ obtained via the FVM. In the present study, a computational domain is defined by intercepting a portion of the whole region in which the cylinder is settling. The geometric description and boundary conditions of the domain are provided in Fig.~\ref{fig:cyl_settling_points_loss}(a). The simulation parameters are set as follows: the gravity $g=1$, the cylinder diameter $D=0.2$, the fluid density $\rho=1$, and the kinematic viscosity $\nu=0.001$. In this computational region, the maximum velocity of the settling cylinder is approximately $U_{max}=1.79$, resulting in a Reynolds number of $Re={U_{max}D}/\nu=358$ for this problem. The Neumann velocity conditions and the zero pressure condition are imposed on the left and right sides of the far-field boundaries, as depicted in Fig.~\ref{fig:cyl_settling_points_loss}(a). We impose Dirichlet velocity conditions on the boundary of the cylinder, where the velocities $u$ and $v$ are determined based on the positions $x$ and $y$ of the cylinders. Notably, we employ the velocities $u$ and $v$ at the grid points of the FVM as labeled data to train the neural network.

\begin{figure}[htb]
    \centering
    \includegraphics[width=1.0\linewidth]{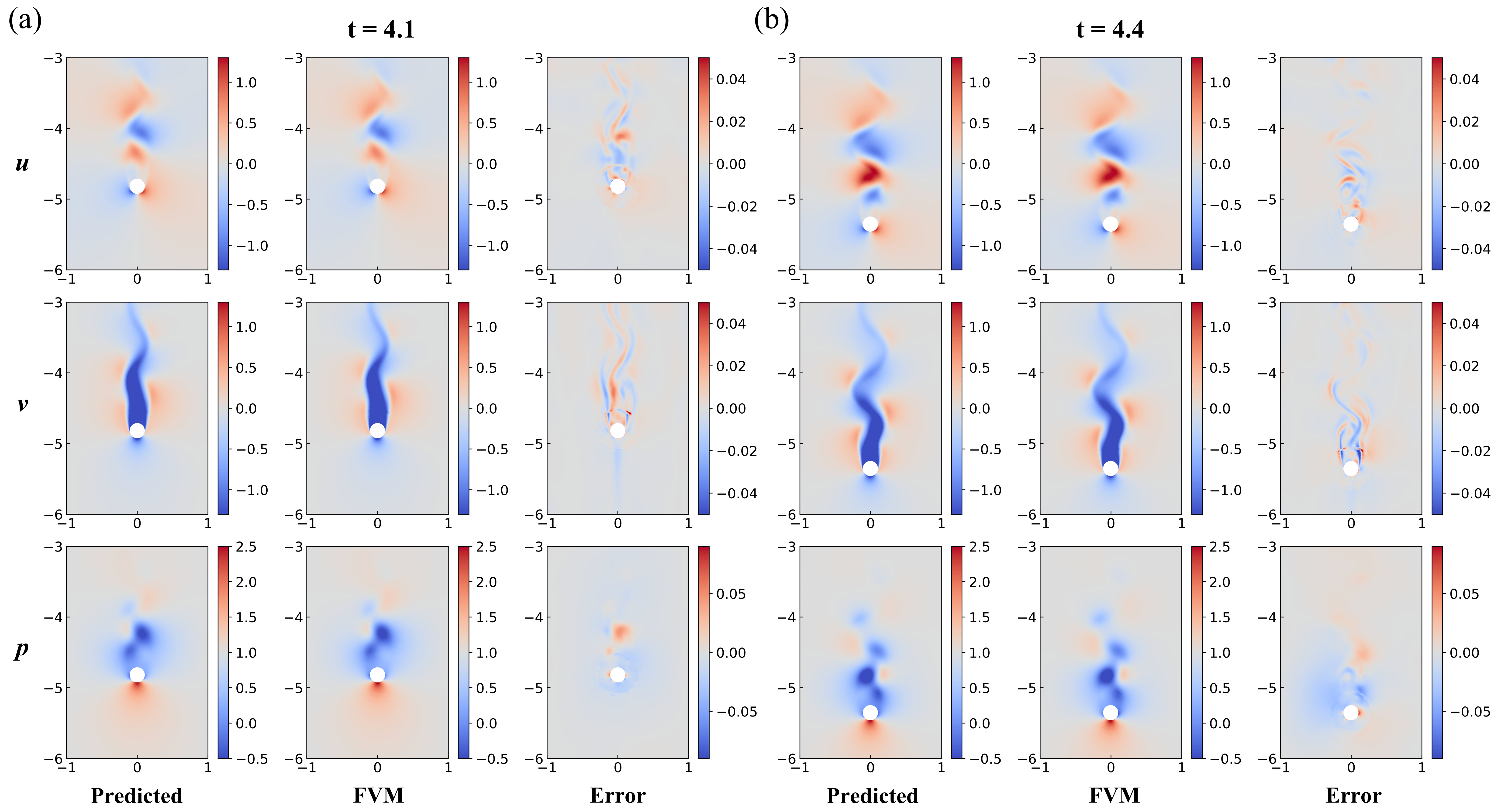}
    \caption{A cylinder settling under gravity: PINNs-inferred velocity $u$, $v$ fields and pressure $p$ field contours at two time frames are in the first columns for (a) $t$=4.1; (b) $t$=4.4. The corresponding results of FVM are in the second columns. The point-wise absolute errors for each frame are in the third columns.}
    \label{fig:cyl_settling_uvp}
\end{figure}

The cylinder settles from rest at $y$ = 0. The selected spatial domain for simulation corresponds to a rectangular region bounded by $[-1, 1] \times [-6, -3]$. And the time domain is [4.0, 4.4]. The location and velocity of the settling cylinder are obtained by fitting the Fourier series to the cylinder trajectory extracted from the FVM results. We excavate the tunnel formed by the settling cylinder in the whole spatial-temporal domain. A visual representation of this tunnel corresponding to the settling trajectory is shown in Fig.~\ref{fig:cyl_settling_points_loss}(b). A total of 229,433 data points, corresponding to 21 time snapshots, scattering in space and time, have been utilized to perform inference of the pressure field. The time interval between two consecutive snapshots is $\Delta t$ = 0.02. Fig.~\ref{fig:cyl_settling_points_loss}(c) depicts a snapshot at $t$=4.4 demonstrating the distribution of these data points. In addition to the 1575 sparse data points in the 21 time snapshots that are at the cylinder boundary, we randomly sample an additional 10,000 data points on cylinder surface as a supplement. Fig.~\ref{fig:cyl_settling_points_loss}(d) provides an aerial perspective of all the points located at the moving cylinder surface, viewed from both the $x$ and $y$ directions.
The convergence of the total loss and its individual components during training, including PDEs loss, moving boundary loss, stationary boundary loss, and velocity $(u, v)$ data loss, are presented in Fig.~\ref{fig:cyl_settling_points_loss}(e). 

\begin{figure}[htb]
    \centering
    \includegraphics[width=1.0\linewidth]{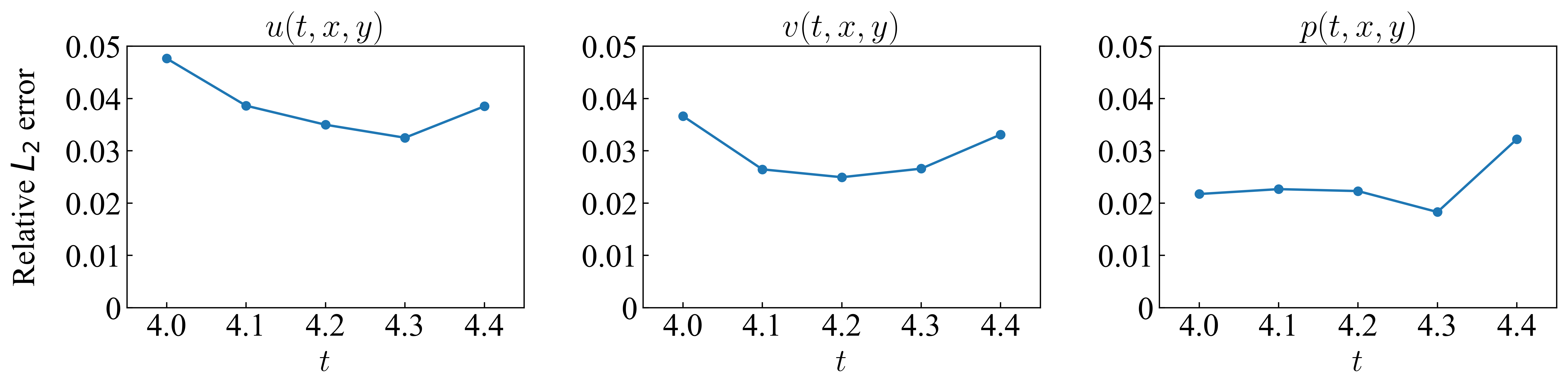}
    \caption{Relative $L_2$ errors of PINNs for a cylinder settling under gravity: the inferred velocity $u$, $v$ and pressure $p$.}
    \label{fig:cyl_settling_error}
\end{figure}

Fig.~\ref{fig:cyl_settling_uvp} shows snapshots of inferred velocity $u$, $v$ and pressure $p$ contours, as well as a visual comparison with the FVM results at two time frames.
The proposed framework is capable of accurately reconstructing the pressure. Fig.~\ref{fig:cyl_settling_error} provides the point-wise relative $L_2$ errors from $t$= 4.0 to 4.4. The network demonstrates the capability to accurately reconstruct the complete flow field. 

\section{Discussions and conclusions}
\label{sec:Discussions and conclusions}
We presented a novel extension to incorporate moving boundary conditions in fluid mechanics into the loss functions of physics-informed neural networks~(PINNs). 
As we interpreted the time-dependent moving boundaries in the spatial domain as stationary boundaries in a higher dimension of the spatial-temporal domain,
we can effortlessly distribute fine training points at and around the interfaces for accuracy.
Therefore, we are able to solve one class of unsteady flow problems with moving bodies,
where the time-dependent interfaces are available via other means.
We have validated the extended PINNs on a variety of classical flow problems involving moving bodies. For instance, we solved the velocity, vorticity, and pressure fields for a complete cycle of an in-line oscillating cylinder in fluid at rest using only data for the initial condition. Additionally, we predicted the flow field surrounding two, three, or four cylinders translating along a circle and demonstrated the effectiveness of this approach for multiple moving bodies. Furthermore, for flows around a flapping wing, we illustrated the flexibility of the proposed framework first by relying data from the initial conditions to predict the later flow field in Section~\ref{sec:Functioning as a DNS solver} and subsequently using data on the velocity $u$ to infer the velocity $v$ and pressure for a reconstruction of the entire flow field in Section~\ref{sec:Reconstructing the flow field}. 
In the case of a transversely oscillating cylinder in flow, we also reproduced the phenomenon of rapid vortex shedding due to oscillation with pressure data.
Lastly, for a cylinder settling under gravity, we relied on both the trajectory of the cylinder and the velocity data to infer the pressure field.
Comparisons with reference solutions for these flow problems demonstrated the effectiveness and accuracy of the extended PINNs.

The extension proposed here may be insightful for flow problems solved by other machine learning frameworks and should be also applicable to physics problems with time-dependent moving boundaries beyond fluid mechanics.
As PINNs exploits a fully connected neural network to approximate the solution of PDEs, currently we cannot achieve the same order of accuracy as the CFD methods within the same wall-time window. 
For example, the in-line oscillating cylinder case took about 3 hours to solve for a full oscillating period on a Nvidia RTX 4090 graphic card.
Improving the accuracy further would require an even larger number of training points
and a more significant amount of epochs for optimization, which both result in a higher computational cost. 
Due to this exact reason, it does not seem necessary to further extend PINNs
to handle the two-way coupling problems,
which may be more feasible by CFD methods currently.
For flow problems with access to the trajectories of moving interfaces,
yet with limited measurements on the flow fields,
the proposed method can be applied steadily.
A potential revolution to the existing neural network architecture and training strategy 
may also fundamentally reform the status and allow PINNs for addressing more computational challenges in fluid mechanics.

\section*{Declaration of competing interest}
The authors declare that they have no known competing financial interests or personal relationships that could have appeared to influence the work reported in this paper.

\section*{Acknowledgments}
Support from the grant of Innovative Research Foundation of Ship General Performance under contract number 31422121 is gratefully acknowledged.
X. Bian also received the starting grant from 100 talents program of Zhejiang University.

\appendix

\section{Technical details of FVM simulations}
\label{Appendix A}
The simulations by FVM in this study have been executed within OpenFOAM. Morphing mesh and overset mesh techniques are constructed for handling the moving boundaries.
We employed the pimpleFoam solver coupled with the morphing mesh technique to solve flow problems in Section~\ref{subsec:In-line oscillating cylinder in fluid},~\ref{subsec:Transversely oscillating cylinder in steady flow}, and used the overPimpleDyMFoam solver coupled with the overset mesh technique to solve flow problems in Section~\ref{subsec:Multiple cylinders translating along a circle},~\ref{subsec:DNS_Flow around a flapping wing},~\ref{subsec:reconst_Flow around a flapping wing},~\ref{subsec:A cylinder settling under gravity}. The time step is automatically adjusted within the limit of the Courant number being 0.4.
Other technical details are given in the captions of the figures as follows.
\begin{figure}[htb!]
    \centering
    \includegraphics[width=0.9\linewidth]{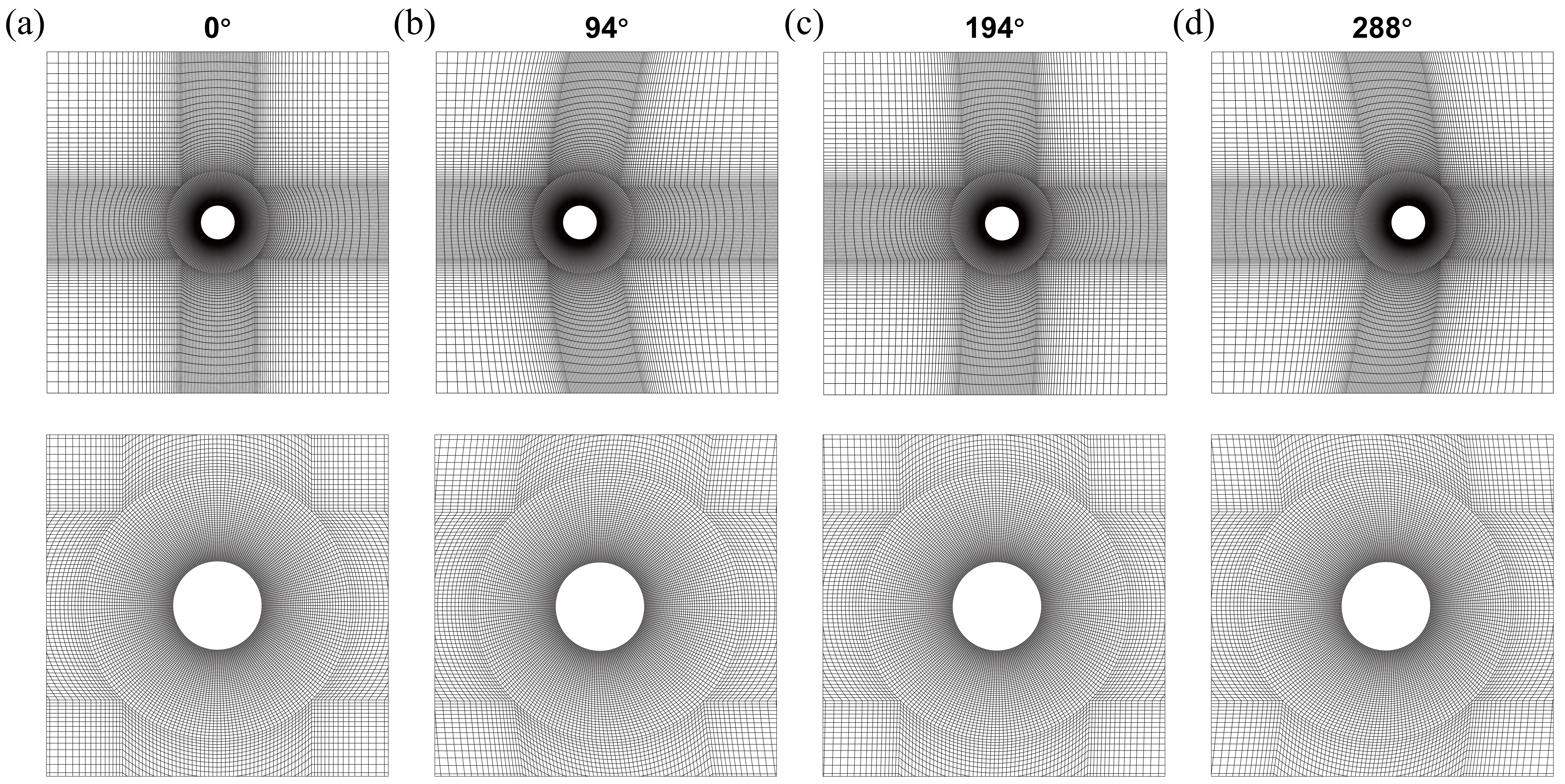}
    \caption{In-line oscillating cylinder in fluid at rest: a morphing mesh with a finer resolution zone of (Diameter=3D) around cylinder is adopted in several typical phases. It contains a total of 19,800 cells and 40,240 nodes. Four distinct phases and corresponding time: (a) 0°(t=0.0); (b) 94°(t=1.3); (c) 194°(t=2.7); (d) 288°(t=4.0).}
    \label{fig:inline_cyl_fvm_meshes}
\end{figure}

\begin{figure}[htb]
    \centering
    \includegraphics[width=1.0\linewidth]{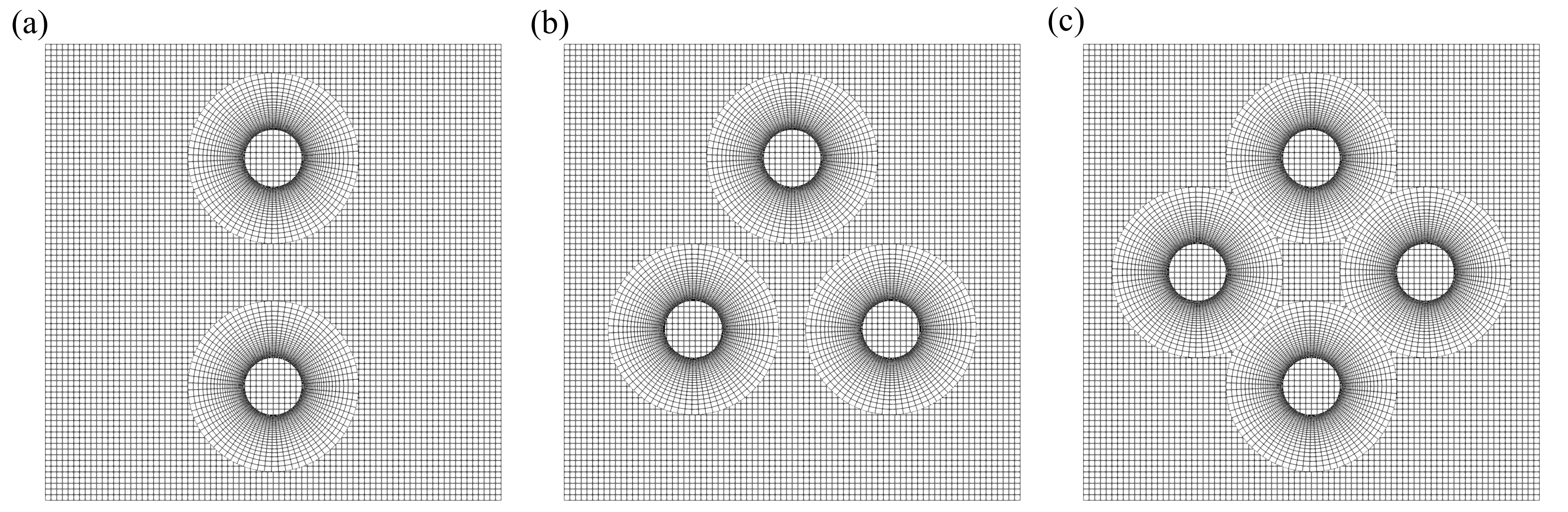}
    \caption{Multiple cylinders translating along a circle: a uniform background mesh (80 $\times$ 80) with finer resolution overset mesh (each contains a total of 2,250 cells and 4,650 nodes) around each cylinder is adopted at time t=2.5. And the left, middle, and right panels depict the configurations of 2, 3, and 4 cylinders, respectively.}
    \label{fig:multi_cyl_fvm_meshes}
\end{figure}

\begin{figure}[htb]
    \centering
    \includegraphics[width=0.9\linewidth]{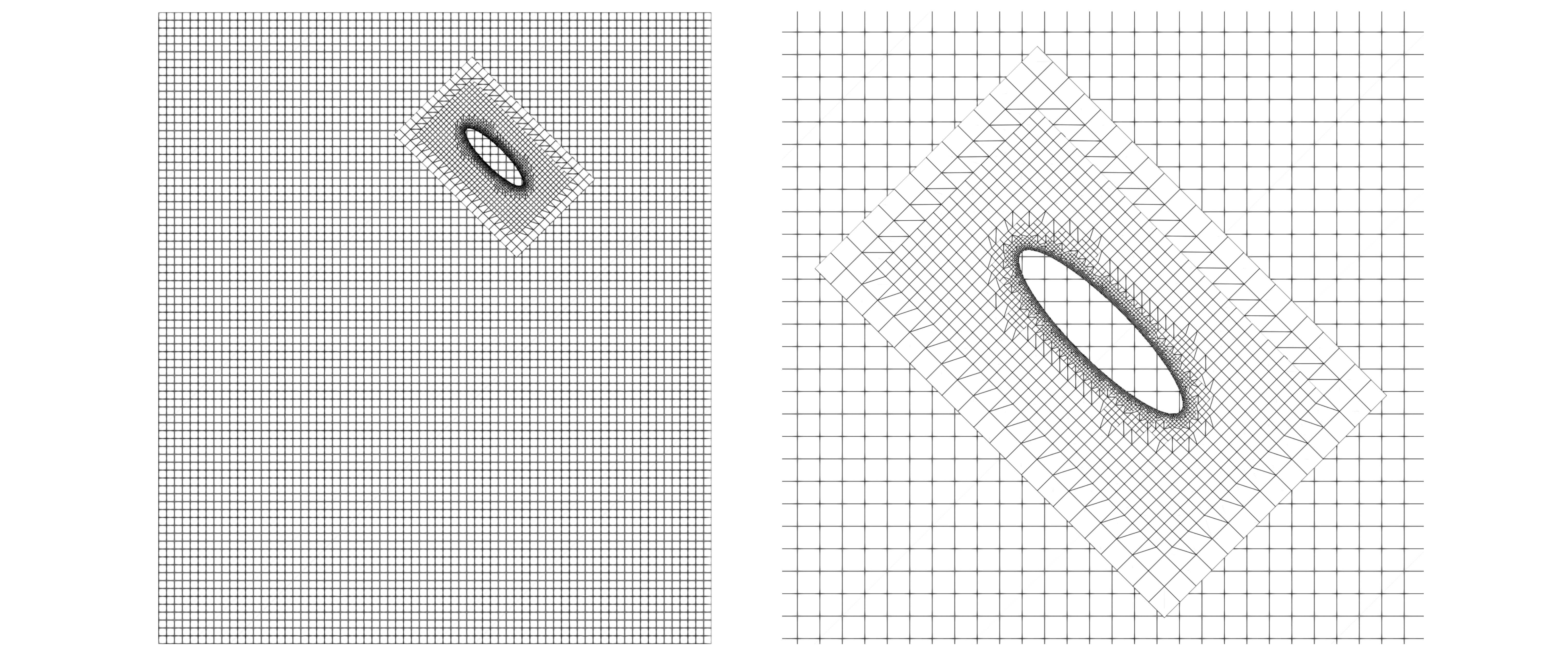}
    \caption{Flow around a flapping wing: a uniform background mesh (70 $\times$ 80) with a finer resolution overset mesh (contains a total of 13,564 cells and 17,720 nodes) around wing is adopted at time t=20. Adaptive mesh refinement technology is adopted to improve the resolution of the mesh near the wing.}
    \label{fig:flapping_wing_fvm_mesh}
\end{figure}

\begin{figure}[htb]
    \centering
    \includegraphics[width=1.0\linewidth]{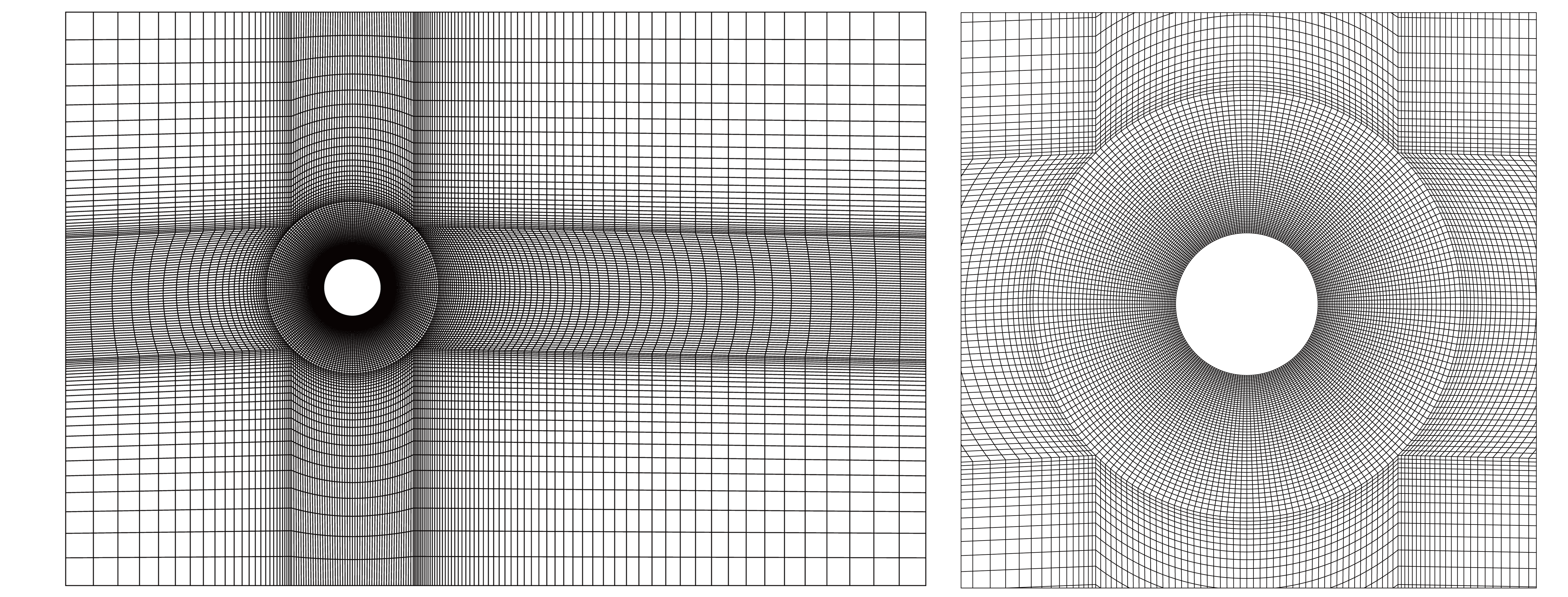}
    \caption{A transversely oscillating cylinder in steady flow: a morphing mesh with a finer resolution zone of (Diameter=3D) around cylinder is adopted at time t=7. It contains a total of 21,200 cells and 43,070 nodes.}
    \label{fig:trans_cyl_fvm_meshes}
\end{figure}

\begin{figure}[htb]
    \centering
    \includegraphics[width=1.0\linewidth]{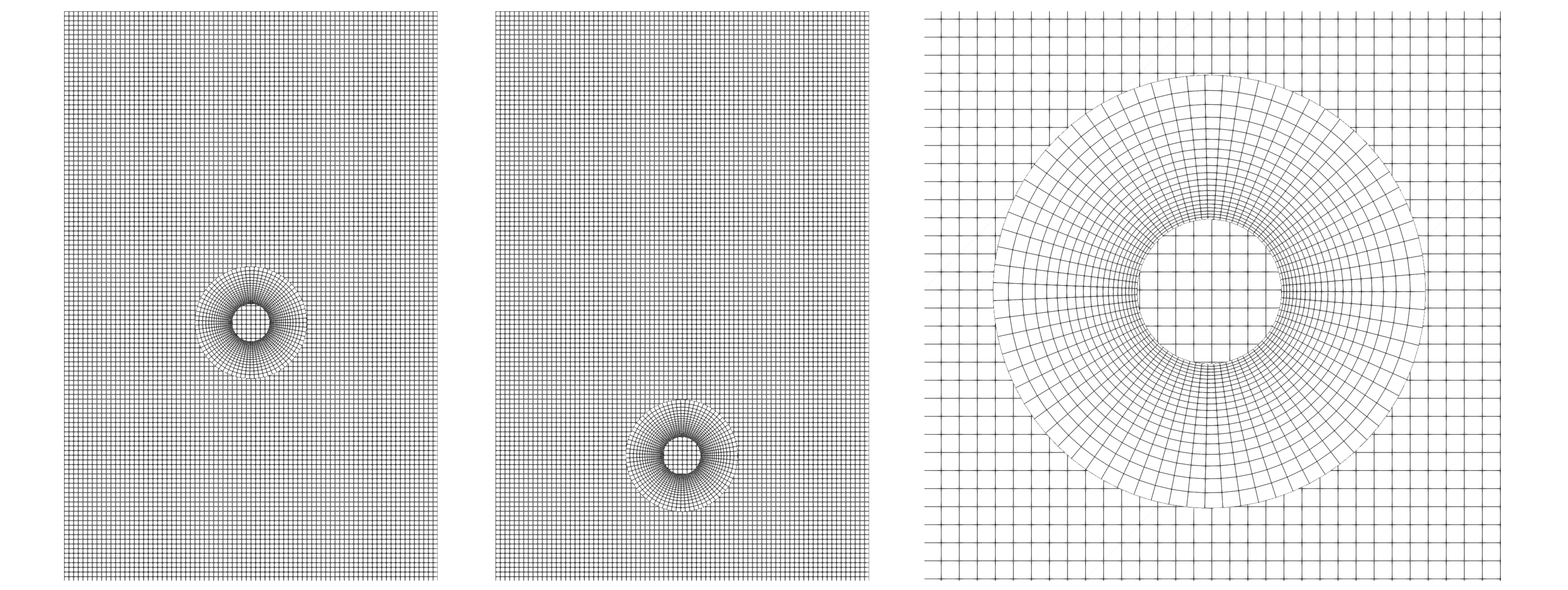}
    \caption{A cylinder settling under gravity: a uniform background mesh (80 $\times$ 120) with a finer resolution overset mesh (contains a total of 2,250 cells and 4,650 nodes) around cylinder is adopted.}
    \label{fig:cyl_settling_fvm_meshes}
\end{figure}


\bibliographystyle{elsarticle-num-names} 
\bibliography{main}





\end{document}